\documentclass[12pt,preprint]{aastex}

\usepackage[normalem]{ulem}
\usepackage{color}

\usepackage{natbib}
\shorttitle{The {\em SIX} survey}
\shortauthors{Bottacini et al.}

\begin{document}

\title{The deep look onto the hard X-ray sky:\\
The {\em Swift} - INTEGRAL X-ray ({\em SIX}) survey}
\slugcomment{\sc Astrophysical Journal Supplement Series; Received 2012 February 7; accepted 2012 June 22}

\author{Eugenio Bottacini\altaffilmark{1}, Marco Ajello\altaffilmark{2} \and Jochen Greiner\altaffilmark{3}}
\email{eugenio.bottacini@stanford.edu}

\altaffiltext{1}{W.W. Hansen Experimental Physics Laboratory \& Kavli Institute
for Particle Astrophysics and Cosmology, Stanford University, USA}
\altaffiltext{2}{Stanford Linear Accelerator Center/KIPAC, 2572 Sand Hill Road, Menlo Park, 
     CA 91125, USA.}
\altaffiltext{3}{Max-Planck-Institut f\"ur Extraterrestrische Physik, Giessenbachstrasse 1, 85748
Garching, Germany}

\begin{abstract}
The super-massive black-holes in the centers of Active Galactic Nuclei
(AGN) are surrounded by obscuring matter that can block the nuclear radiation.
Depending on the amount of blocked radiation, the flux from the AGN can
be too faint to be detected by currently flying hard X-ray (above 15 keV)
missions. At these energies only $\sim$1\% of the intensity of the Cosmic X-ray Background (CXB)
can be resolved into point-like sources that are AGNs. In this work we
address the question of the undetected sources contributing to the CXB
with a very sensitive and new hard X--ray survey: the {\em SIX} survey that is
obtained with the new approach of combining the {\em Swift}/BAT and INTEGRAL/IBIS
X--ray observations. We merge the observations of both missions. This enhances
the exposure time and reduces systematic uncertainties. As a result we obtain
a new survey over a wide sky area of 6200 deg$^{2}$ that is more sensitive than the
surveys of  {\em Swift}/BAT or INTEGRAL/IBIS alone. Our sample comprises 113
sources: 86 AGNs (Seyfert-like and blazars), 5 galaxies, 2 clusters of galaxies,
3 Galactic sources, 3 previously detected unidentified X-ray sources, and 14
unidentified sources. The scientific outcome
from the study of the sample has been properly addressed to study the evolution
of AGNs at redshift below 0.4. We do not find any evolution using the 1/V$_{max}$
method. Our sample of faint sources are suitable targets for the new generation
hard X-ray telescopes with focusing techniques.

\end{abstract}


\section{Introduction}
In the view of the so-called AGN unified model \citep{antonucci93,urry95} a super-massive black hole (SMBH)
harbored at the center of the AGN powers the nuclear activity. The region where the activity takes place
can be observed from different viewing angles. Therefore depending on the orientation of the AGN the observer's line
of sight intercepts different amounts of the optically thick gas--dust structure (torus) that surrounds the SMBH.
The nuclear radiation at optical/UV and X-ray wavelengths is efficiently
absorbed by the torus. The amount of obscuring matter (N$_{H}$ column density associated to the torus)
can be best inferred by X-ray spectra of the AGNs. X-ray surveys are therefore powerful tools for AGN
population studies.
The bias of X-ray surveys strongly depends on the column density associated to the sources and the survey
sensitivity: the larger the column density and the worse the flux sensitivity, the better the low--absorbed
sources are selected. Such selection effect is negligible for unabsorbed sources
(exhibiting N$_{H}$ $<$ 10$^{22}$ cm$^{-2}$) while it affects the absorbed sources
(exhibiting N$_{H}$ $>$ 10$^{22}$ cm$^{-2}$) and it is magnified for sources with column
densities N$_{H}$ $\ge$ 1.5 $\times$ 10$^{24}$ cm$^{-2}$. This latter value corresponds to the inverse of the Thompson
cross-section ($\sigma_{T}^{-1}$) and the optical depth unity for Compton scattering. Absorbed sources
affected by such high column densities are defined as "Compton-thick". This plays an important
role in nowadays most sensitive AGN X-ray surveys that are performed by {\em XMM-Newton} and {\em Chandra}
in the energy range $\sim$0.5 - 10 keV \citep{brandt01,alexander03,cappelluti09,xue11}. 
At these energies less than a mere 10\% of the nuclear radiation is energetic enough
to pierce through the absorbing Compton-thick torus \citep{gilli07}. On the other hand the efficiently
absorbed optical/UV radiation heats the dust of the obscuring medium, that is expected to waste the absorbed
radiation in form of IR emission. Indeed, an IR--excess due to warm dust heated by obscured AGNs has
been found \citep{fadda02}. Infrared power-law selected samples in {\em Chandra} Deep Fields are promising
AGN--candidates \citep{alonso-herrero06,donley07}. The drawback of the IR selection is that the majority
of the detected sources are not AGNs. Furthermore this approach seems to sample best the sources within
redshift 1--3 \citep{donley07}. This is the same redshift range in which {\em Chandra} and {\em XMM-Newton} are preferentially selecting most AGNs in their deep surveys \citep{brandt-hasinger05,hasinger08}.
Instead the redshift space at z $<$ 0.4 is so far poorly explored despite extensive studies \citep{markwardt05,beckmann06,sazonov07,ajello08c,tueller08,bird10,cusumano10}.

The low-redshift (z $<$ 0.4) Universe is best fathomed at hard X--ray energies ($>$ 15 keV).
With the advent of the INTEGRAL \citep{winkler03} and the {\it Swift} \citep{gehrels04} missions,
the selection of local AGNs through their hard X--ray ($>$15 keV) emission has proven to be
an extremely powerful technique over the last few years.
INTEGRAL and {\it Swift} carry coded-mask telescopes on board, namely the Imager on--Board 
the INTEGRAL Satellite \citep[IBIS:][]{ubertini03} and the Burst Alert Telescope
\citep[BAT:][]{barthelmy05} respectively. IBIS has two detector layers. One of which is the
INTEGRAL Soft Gamma-Ray Imager \citep[ISGRI:][]{lebrun03}. 
IBIS/ISGRI and BAT have two major advantages:
1) they have a huge field of view, hence allowing to sample an adequate number of AGNs at low-redshift
2) they operate at energies above 15 keV, hence allowing detecting the photons having enough penetrating 
power to pierce efficiently even through the Compton-thick torus. A further and major advantage in 
sampling photons above 15 keV from AGNs comes from the emitting source itself. Indeed, a broad 
continuum bump, so-called "Compton-reflection bump", peaking at energies between 20 - 30 keV is 
produced by reflection of the primary nuclear radiation on the inner side of the obscuring 
gas \citep{george91,gilli07}. This spectral component has been found to be dominant
in nearby heavily  obscured AGNs \citep{comastri07}. The Compton-reflection component also plays
an important role in reproducing the shape and intensity of the CXB \citep{rogers91,gilli01,ueda03},
that peaks at 30 keV \citep[for most recent measurements see:][]{ajello08c,moretti09,tuerler10}. 
Estimates based on observations with PDS \citep{frontera97} on board the {\it Beppo}SAX 
\citep{boella97} satellite predict that Compton-thick AGNs should dominate over unobscured AGNs 
in the local Universe \citep{matt00}.
This makes IBIS/ISGRI and BAT well suited instruments for detecting obscured AGNs in the local Universe. 
IBIS/ISGRI and BAT both represent a major improvement for the imaging of the sky above 15 keV. 
However coded-mask detectors suffer from heavy systematic effects (errors) preventing them from 
reaching their theoretical limiting sensitivity \citep{skinner08}. Furthermore by design they 
block $\sim$50\% of the incident photons causing an increase of the statistical noise. 
These are the reasons that make the detection of extragalactic sources,
that are mostly faint, still challenging to undertake. Here
we describe an alternative approach which has been developed
ad hoc to improve the sensitivity of extragalactic hard X--ray
surveys by using IBIS/ISGRI and BAT.

In this paper we show that {\it Swift}/BAT and INTEGRAL/IBIS observations can be merged
to obtain a more sensitive survey that is able to sample limiting fluxes of 
$\sim$3.3 $\times$ 10$^{-12}$ erg cm$^{-2}$ s$^{-1}$ in the 18 - 55 keV energy range.
We call this the {\em SIX} survey, that stands for {\it Swift}--INTEGRAL hard X-ray survey.
The {\em SIX} survey extends over a wide sky area of 6200 deg$^{2}$ and it is used to 
obtain a small and persistent sample of faint sources. This enables the construction of the 
number density (log $N$--log $S$) as well as developing the X-ray luminosity function (XLF) for AGNs. 
In addition we estimate the contribution of this sample of AGNs to the intensity of the 
unresolved fraction of the CXB. Throughout this paper we adopt the cosmological parameters 
of: H$_{0}$ = 70 h$_{70}$ km s$^{-1}$ Mpc$^{-1}$, $\Omega_{\Lambda}$ = 0.73.

\section{The {\em SIX} survey}
IBIS/ISGRI and BAT are both coded--mask instruments. Their performances are considered as a
milestone for the sky imaging at hard X-ray energies. They pose themselves as 
excellent instruments for population studies of faint hard X-ray sources shedding
continuously light on the properties of the local AGN population \citep{beckmann06,ajello08a,ajello09b,
tueller09,bird10,cusumano10}.
Currently the two instruments show a difference in the extragalactic 
sky survey. This is mainly due to the different pointing 
strategies adopted for the satellites. {\it Swift} is quasi-randomly
pointing the sky, while INTEGRAL performs targeted observations and
long exposures on the Galactic Plane. The authors of the 4$^{th}$ IBIS/ISGRI catalog
\citep{bird10} conclude that the non--detection of BAT--detected
sources by IBIS/ISGRI is just due to the low exposure of those sources
in the ISGRI detector. Indeed, at comparable exposure time on the extragalactic
sky the BAT sample \citep{cusumano10} contains $\sim$70\%
extra-galactic sources, the IBIS/ISGRI sample contains
$\sim$35\% \citep{bird10}. Table~\ref{tab:bat-isgri} summarizes the 
in--flight performances of the two instruments. For IBIS/ISGRI the flux
sensitivity in the 20--40 keV energy range at 4.8$\sigma$ is computed
over 90\% of the extragalactic sky \citep{bird10}. The BAT flux
sensitivity is obtained over the entire extragalactic sky in the
15--30 keV energy range at 4.8$\sigma$ \citep{cusumano10}.\\
\begin{table}
\begin{center}
\caption{Comparison of BAT and IBIS/ISGRI in-flight Performances.\label{tab:bat-isgri}}
\begin{tabular}{lccccc}
\hline\hline
Parameter & IBIS/ISGRI & BAT \\
\hline
PSF (arcmin) & 12 &  22 \\
FOV (deg$^2$) & 400 &  4500 \\
Energy range (keV) & 17--1000 &  13--300 \\
\hline
\end{tabular}
\end{center}
\end{table}
To obtain the {\em SIX} survey, we first perform the independent surveys
of  BAT and IBIS/ISGRI. Then by combining the observations of the two instruments we
increase (sum) the exposure time. In turn the sensitivity of the {\em SIX} survey is enhanced.
We compute the survey over a sky area of 6200 deg$^{2}$ that covers the region
of North Ecliptic Pole (NEP) extending to the contiguous Coma region. We have
chosen this sky area because it is covered to a large exposure time by both, BAT
and IBIS/ISGRI. In addition,
ROSAT has covered this area to a deep sensitivity \citep{voges99} making the
identification of the {\em SIX} sources robust.
We perform the {\em SIX} survey in the 18--55 keV  energy range even though BAT
is sensitive to 200 keV and IBIS/ISGRI to even higher energies (1 MeV).
The lower limit is due to the physical energy threshold of both detectors.
The upper energy limit is related to sensitivity issues. Since, our aim is
to perform a very sensitive hard X-ray survey, we try to avoid  systematic
effects due to background lines and possible uncertainties in the instruments' 
response. Furthermore, we want to take advantage of the Compton--reflection 
bump in AGN spectra peaking in the 20--30 keV range. The contribution of this
spectral feature decreases rapidly at high energies because of Compton
down--scattering. Therefore we set the upper threshold to 55 keV.
\subsection{Analysis of BAT data}
BAT is a coded mask telescope with a wide field of view 
(FOV, $120^{\circ}  \times 90^{\circ}$ partially coded) aperture
sensitive in the 15--200\,keV domain.
BAT's  main purpose is to locate Gamma-Ray Bursts (GRBs).
While  chasing new GRBs, 
BAT surveys the hard X-ray sky with an unprecedented sensitivity.
Thanks to its wide FOV and its pointing strategy, 
BAT monitors continuously up to 80\% of the sky every day.
Results of the BAT survey \citep{markwardt05,ajello08b} 
show that BAT  reaches a sensitivity of $\sim$1\,mCrab in 1\,Ms of exposure.
Given its sensitivity and the large exposure already accumulated in the whole
sky, BAT poses itself as an excellent instrument for looking 
for the (faint) emission of AGNs above 15\,keV.\\
For the analysis presented here, we used all the available data
taken from March 2005 to March 2010. 
The analysis method and software
are described in \cite{ajello08b}. The analyzed energy interval
ranges from 18--55 keV as explained in \S2.3. 
The data screening was performed according to \cite{ajello08b}.
The all-sky image is obtained as  
the weighted average of all the shorter observations.
The average exposure time in our image is 3\,Ms, being 1.3\,Ms and 5\,Ms the minimum
and maximum exposure times respectively. 
\subsection{Analysis of IBIS data}
The IBIS imager on--Board the INTEGRAL Satellite
is a coded--mask instrument \citep{goldwurm01} for the imaging of the
sky in the energy range 15 keV -- 10 MeV. ISGRI
is the low--energy detector array of IBIS in the domain 15--1000 keV
with a wide field of view (FOV, 29 x 29 deg$^2$) and an angular
resolution of 12$\arcmin$ (FWHM). IBIS hosts a further detector layer
(PICsIT: Pixellated CsI Telescope) operating at energies
175 keV -- 10 MeV \citep{dicocco03}. The main goal of IBIS is to study
point--like sources.
Thanks to its FOV and while pointing predetermined coordinates, IBIS/ISGRI
is monitoring large areas of the sky. We reduced the data according to the standard 
Off-line Scientific Analysis (OSA) software version 7.0
\footnote{http://www.isdc.unige.ch/integral/download/osa\_doc}
 \citep{courvoisier03}.
OSA is based on cross-correlation method \citep{goldwurm03}.
In addition we apply an iterative source removal for image
reconstruction. Due to the cyclic mask pattern of IBIS coded-mask, 
the OSA software does not completely remove ghosts, caused by
bright and/or extended sources, in specific positions
of sky. This adds not negligible systematic
errors to mosaic images having long exposures. Therefore, particular
attention was drawn on this issue detailed in the following.
\subsubsection{Analysis description}
The aim of this analysis is to obtain IBIS/ISGRI sky images whose
systematic uncertainties are reduced. Our dataset covers INTEGRAL
pointings (that are Science Windows - ScWs in the following) from the 
region centered around the Coma sky--area and extending to and including
the North Ecliptic Pole (NEP). 
On this sky area we have used all public available data as well as private
data (PI M. Ajello, proposal ID: 05K001). The data set spans over 7 years from
the beginning of the mission (year 2002) to INTEGRAL revolution 829 (year 2009). 
The total exposure is $\sim$12 Msec. Most pointings are performed
following a $5 \times 5$ dithering pattern. We have checked that no
staring observations are included that would complicate our analysis.
The pointed observing strategy and the dither pattern adopted by INTEGRAL
yield a non--uniform exposure.
IBIS/ISGRI data come in form of photon-by-photon basis, meaning that
each event in the detector is tagged according to the detector coordinates,
event energy deposit and event time. Each ScW has a typical exposure
time between 2500 and 3000 seconds. Particular attention has been
drawn to those lasting for longer and shorter time scales possibly affected
by perigee passage of the satellite or other issues.
\subsubsection{IBIS/ISGRI sky maps}
OSA produces a sky image of each single ScW. The software corrects
for noisy pixels and converts the channels to energy accounting for
photon rise-time and gain variability. Also the dead-time is accounted
for. For the chosen energy range an intensity shadowgram (detector plan) and
efficiency pixel map is computed. Pixels are corrected for efficiency
and the background map, which is derived from flat--field observations,
is subtracted. A source in the FOV projects a mask pattern onto the
pixellated detector plane. This is known as coding phase. A decoding
phase is required that allows to reconstruct the original sky. Therefore
a mask pattern is used as a deconvolution array applying fast Fourier
transforms (cross-correlation). Once each ScW has been de--convolved 
OSA produces a mosaic image of the observed sky region.
\subsubsection{Subtraction of bright sources}
Unlike for conventional focusing telescopes, the Point Spread Function
(PSF) of coded-mask detectors extends over the whole detector plan.
The consequence is that the PSF of each bright source in the FOV
introduces fluctuations in the de-convolved image that can exceed
the statistical noise.\\ 
The only way forward to account for the fluctuations introduced by
the bright source's PSF over the whole detector shadowgram is to compute a 
shadowgram pattern of the mask that is cast by the source onto the
detector plane. The intensity of this pattern can be fit and then be
subtracted in the detector space avoiding the PSF to affect the quality of
the final reconstructed sky image. The mask pattern for each source in the FOV 
and for each single ScW can be easily computed by OSA itself and it is called
Pixel Illumination Function (PIF). We compute a PIF for each source as
it is detected in the FOV for each ScW. We fit simultaneously the set of PIFs
to the detector intensity map. Therefore we can compute the intensity of each single
source simultaneously in each single ScW. The simultaneous fit allows us to account
naturally for the variability of the source intensity as it is observed
by the instrument. The resulting cleaned detector shadowgrams are used
to construct with OSA the new ScW--images.
Due to the lower noise level, weaker sources can arise in the new mosaic.
Furthermore, bright sources are intrinsically variable. When performing
a survey with long exposure time, these sources can pop--up just for a
short time, remaining hidden (below the significance detection threshold)
in the survey mosaic image. Therefore they contribute to the background
level in the mosaic image. An explanatory example for this is given by
1ES~1959+650. This BL Lac blazar is located in our surveyed NEP area.
The source is detected by IBIS/ISGRI within $\sim$4 ksec \citep{bottacini10}
while it is not detected when integrating over the whole exposure time
on the NEP as it is not included in the 4$^{th}$ IBIS/ISGRI catalog \citep{bird10}.  
\subsubsection{Modeling the background}
There is a difference between the background used by the cross-correlation
algorithm  and the background of real coded-mask instruments \citep{skinner08}.
This latter background faces a non-spatial uniformity in the detector plane
(due to instrumental noise and cosmic environment) that can be addressed by OSA
itself by means of re-normalized and balanced cross-correlation \citep{goldwurm03}.
The variability of the background intensity is on time scales of hours to days.
Variability on longer time scales (months to years) is due to solar modulation 
and changes of the instrument performances. This additional variable background
component is not addressed by OSA. After subtracting the background models
provided by OSA and our bright source models we obtain an improved detector
plane shadowgram. We use these shadowgrams to compute residual maps on time
scales of 6 months. Each residual map is calculated by weighting for the exposure
of the single ScW.
\subsubsection{Final image reconstruction}
After computing the bright source models and the residual maps we use them
as models and we fit them at once to the original detector shadowgrams (for each ScW).
The fitting algorithm performs the solution using a Cholesky
decomposition scheme for solving a system of linear equations. The fitted
components are then subtracted from the data of the detector plane shadowgram.
The new shadowgram is finally used by OSA to reconstruct the sky.
We finally test our background model on ScW level. After subtracting the modeled
background we find that the detector pixel distribution is Gaussian. Less than
$\sim$1\% is outside the 3$\sigma$ confidence level. This indicates that a number
of pixels are not correctly modeled which can be attributed to not perfect PIF models.
This is a known issue for coded--mask detectors.
\subsubsection{Data selection and data screening of IBIS data}
In order to build the best possible IBIS/ISGRI mosaic image we used
a suitable clean dataset of ScWs as input for the image reconstruction. 
To this end a list of bad time intervals accounting for detector anomalies
(isgr\_gnrl\_bti\_0004.fits) is provided with the OSA.
ScWs exhibiting background fluctuations larger than 1.1 in the significance
maps are rejected. Data screening was performed according to the median
count rate with respect to each ScW and their distribution.
Our data quality cut rejected 1.5\% of the analyzed ScWs. All these ScWs
are characterized by the perigee passage (entering or leaving) of the
satellite where the detector background cannot be modeled due to interaction
of cosmic rays. 
\subsection{Combining the IBIS/ISGRI and BAT mosaic images}
\begin{figure}[ht!]
\begin{center}
 \includegraphics[scale=0.75]{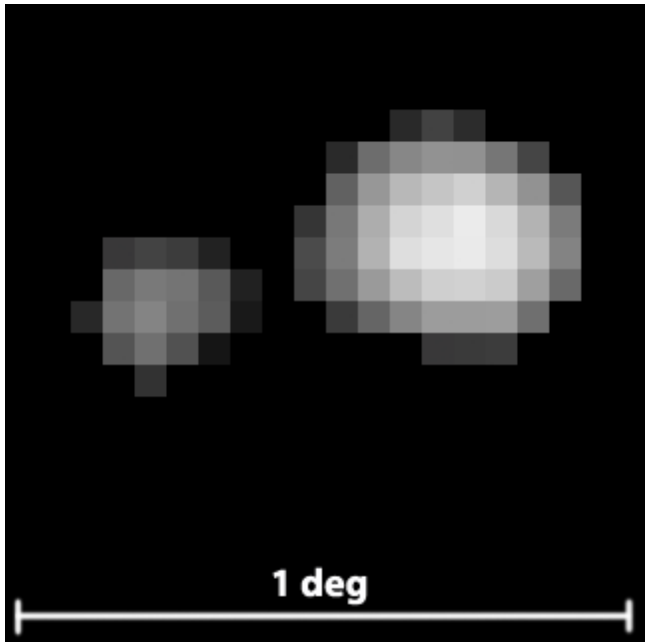}
 \end{center}
\caption{Image of 2E~1923.7+5037 and V* CH Cyg, left
and right respectively. The sources are clearly separated by an angular distance of
29$\arcmin$. The sky--region is $\sim$1 $\times$ 1 deg$^{2}$ and the pixel size is 0.04 deg. 
\label{fig:source-separation}}
\end{figure}

At this point of the survey the independently obtained mosaic images by BAT
and IBIS/ISGRI in the same energy range (18--55 keV) are ready to be merged.
At first the BAT and IBIS/ISGRI sky maps must be re-sampled. After that a
cross-calibration of the maps is performed. Finally we compute the
significance maps, where the sources can be searched. The single steps
are out-lined in the following.\\
The image resampling consists in determining the intensity at an arbitrary
point ($x$,$y$) starting from known image intensities at discrete values
($u$,$v$) in a coordinate system ($X$,$Y$). The approach is to fit a surface
to the discrete points and estimate the surface value at ($x$,$y$). The fastest
method in terms of computing time is to set the intensity in ($x$,$y$) to the
nearest-neighbor value [round($x$),round($y$)]. The drawback of this method is
the aliasing effect (intensities are duplicated or lost) along the edges of 
the mosaic image. In the case of not uniform exposure of the mosaic image
(as for the IBIS/ISGRI sky coverage) this would result in heavy systematics at
the edges of the mosaic. We therefore derive the intensity in the sample
location by interpolating quadratically the nearest neighbor intensities.
This method consists in choosing the 4 nearest pixel values surrounding
the position of the pixel whose value has to be determined. The final
interpolated intensity value is obtained weighting for the pixel's distance.
Suppose that we want to estimate the value of the unknown intensity $I$
at the point ($x$,$y$). The matrix notation of the inferred $I(x,y)$ is:
\begin{eqnarray}
I(x,y) = 
 \left[
 \begin{array}{cc}
     1 - x & x - u
 \end{array} 
 \right]
 \left[
 \begin{array}{cc}
     I(u,v) & I(u,v+1) \\
     I(u+1,v) & I(u+1,v+1) \\
 \end{array} 
 \right] 
 \left[
 \begin{array}{c}
     1 - y \\
     y - v
 \end{array} 
 \right]   
\end{eqnarray}
where $I$($u$,$v$), $I$($u$,$v+1$), $I$($u+1$,$v$) and $I$($u+1$,$v+1$)
are known. The result of this interpolation is independent of the order
of interpolation itself.\\
We re-sample the intensity and the variance maps of both, IBIS/ISGRI and BAT.
IBIS/ISGRI produces sky maps that have a finer angular resolution than the one
of BAT. Therefore we overlay to the coarse BAT intensity sky image a
grid matching the IBIS/ISGRI angular characteristics. Therefore we compute BAT sky
images having pixel size of $\sim$2.4$\arcmin$.\\
After the re-sampling we cross-calibrate the maps of IBIS/ISGRI and BAT.
For both surveys we use the Crab counts spectrum F(E)
in units of [{\rm photon cm$^{-2}$ sec$^{-1}$ keV$^{-1}$}] to determine the
Crab flux (see eqs.~\ref{eq-crab}) and to perform the cross-calibration. For each
pixel of the IBIS/ISGRI map and the BAT map the count rate is cross-calibrated.
At this point the maps are re-sampled, meaning that the maps are aligned along the
same direction and they are of the same pixel size, and cross-calibrated.
The merging of the intensity maps consists in simply summing the maps by
weighting for the errors. The variance maps instead are merged using the following
formula for the error propagation:
\begin{equation}
\sigma_{SIX}^{2} = \sigma_{BAT}^{2} + \sigma_{ISGRI}^{2} + 2\times\textup{cov$_{BAT,ISGRI}$}
\label{eq-error}
\end{equation} 
where $\sigma_{BAT}^{2}$ and $\sigma_{ISGRI}^{2}$ are the variance
terms for each single pixel of BAT and ISGRI respectively. The covariance
term cov$_{BAT,ISGRI}$ $= 0$ being the systematic errors associated to the
respective instruments not correlated. We divide the newly computed intensity
map by the newly calculated noise map obtaining the significance mosaic
of the {\em SIX} survey. We show the capability of our approach
to reconstruct the sky in Figure~\ref{fig:source-separation}. This figure shows
that the 2 closest sources in the survey are clearly separated.

\section{The {\em SIX} Survey: Results}
\subsection{Performance}
\subsubsection{Mosaic properties}
\begin{figure}[ht!]
\begin{center}
 \includegraphics[scale=0.75]{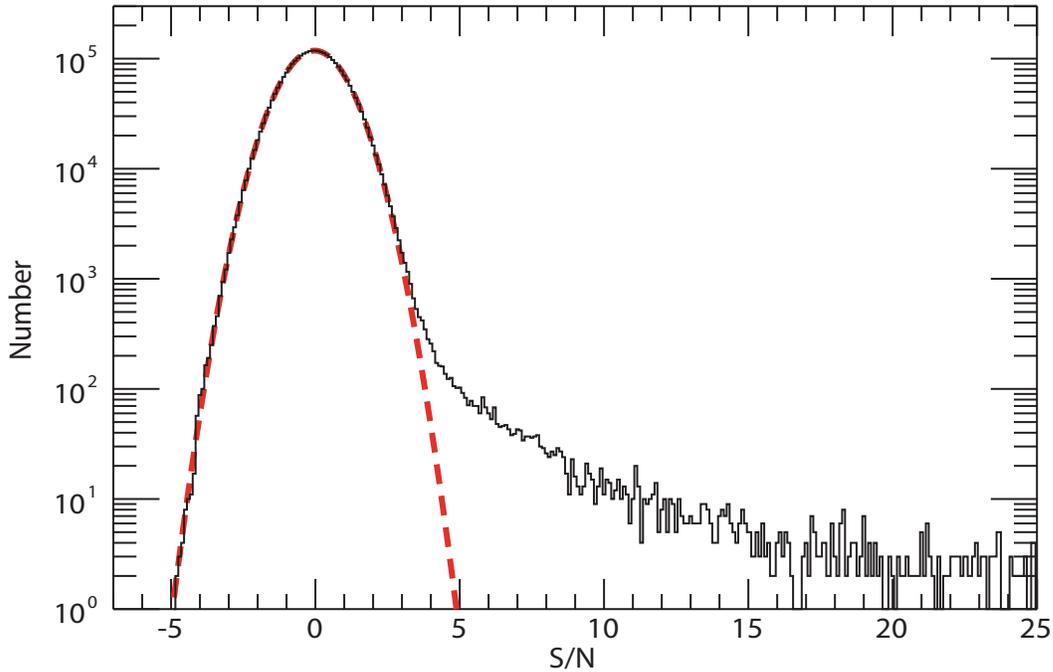}
 \end{center}
\caption{Pixel significance distribution of the {\em SIX} mosaic. The dashed line is
an overlaid Gaussian with standard deviation $\sigma$ = 1.0. The distribution
does not show any wings and the long tail at positive significances are real
detected sources. (A color version of this figure is available in the online journal.)
\label{fig:pixel-distribution}}
\end{figure}

The {\em SIX} survey covers 6200 deg$^2$. To study the quality of the
mosaic image, we investigate its pixel distribution. The distribution
is represented by the black solid line in Figure~\ref{fig:pixel-distribution}
where the red line is an overlaid Gaussian. Its mean value is 0 while the dispersion
$\sigma$ = 1.0. At negative significances no wings are present. The long tail at positive significances
represents real detected sources. The pixel-significance distribution
demonstrates the quality of the background modeling.

\subsubsection{Detection threshold}
To identify an excess caused by a source in a mosaic image it is necessary
to define the significance level at which the source population dominates
over the noise distribution. To do so, we study the
distribution of the pixel significances. In Figure~\ref{fig:pixel-distribution}
the largest negative fluctuation in the pixel distribution is found at $S/N = -4.8$.
Taking into account the Gaussian distribution, we compute the number of
pixels having the S/N-value above 4.8$\sigma$ not caused by the contribution
of the sources but only due to statistical fluctuation. This is 
done by calculating the complementary error function. The value obtained
is then multiplied by 0.5 since only positive fluctuations
(the distribution's tail at positive significances) can give rise
to false detections. We then  multiply this probability with the
number of pixels ($\approx$ 3 $\times$ 10$^{6}$). We find
that only 2 pixels exceed the 4.8$\sigma$ detection threshold by chance.\\
The source search algorithm is based on the {\em Swift}/BAT standard tool
{\it batcelldetect}. It uses the sliding cell method that detects a source at the
position in the image where the signal of a pixel exceeds the background by
our chosen detection threshold. However the oversampling
of the BAT mosaic image might contribute to the fact that the detection threshold
is exceeded by chance by the 2 pixels in the {\em SIX} mosaic. To avoid detecting such 
fluctuations as spurious sources, we require that at least 4 contiguous pixels
exceed the detection threshold. Therefore, by setting the detection threshold to
4.8$\sigma$ we do not expect any false detection.\\
\subsubsection{The {\em SIX} point spread function}
\begin{figure}[ht!]
\begin{center}
 \includegraphics[scale=0.75]{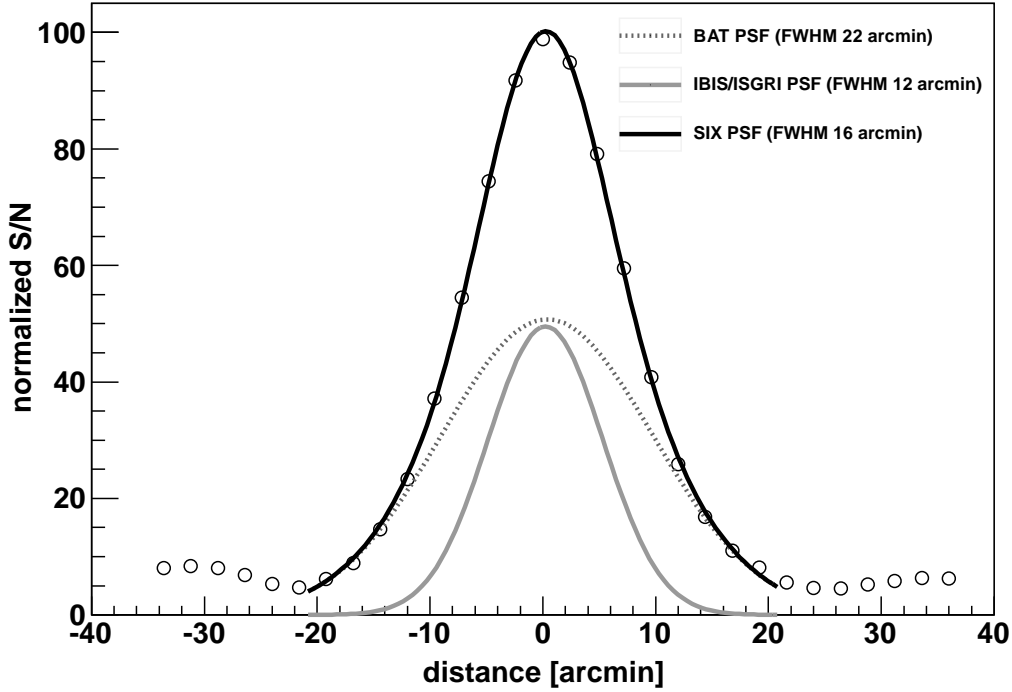}
 \end{center}
\caption{The {\em SIX} PSF (solid black line and FWHM 16$\arcmin$) is a linear
combination of the BAT PSF (dotted gray line and FWHM 22$\arcmin$) and IBIS/ISGRI
PSF (solid gray line and FWHM 12$\arcmin$). 
\label{fig:psf}}
\end{figure}

We fit the {\em SIX} PSF to the region where the sources are detected. This results
in accurate localization of the centroids of the sources. To determine the {\em SIX}
PSF we have extracted the PSF of a large number of {\em SIX} sources without
any preferred direction. This allows us to test the symmetry of the shape. The single
source PSFs were normalized so that we can compute an overall mean PSF.
The data points of the mean PSF are plotted in Figure~\ref{fig:psf} using open circles. The
PSF profile can be modeled by a linear combination of the BAT PSF (dotted line) and the
IBIS/ISGRI PSF (gray line). The BAT PSF
\citep{markwardt05} and the IBIS/ISGRI PSF \citep{gros03} both have Gaussian
shapes with standard deviations of 9.4$\arcmin$ and 5.1$\arcmin$ respectively. These values have
been held fixed to perform a fit that minimizes $\chi$--square with $MINUIT$ \citep{james75}.
The normalization and mean values of both
gaussians were free to vary. The resulting shape (black solid line in
Figure~\ref{fig:psf}) of the {\em SIX} PSF is symmetric with standard deviation of 6.8$\arcmin$.
The derivation of the {\em SIX} PSF from linear combination
of the PSFs of the single instruments resembles the method used
to obtain the {\em SIX} mosaic image that is a linear combination of
the sky maps of BAT and IBIS/ISGRI.

\subsubsection{Source flux}
We extract from the {\em SIX} intensity map the fluxes of detected sources.
The flux is computed by converting the count rate (cts) to
physical units [erg cm$^{-2}$ s$^{-1}$] making use of the Crab as a
calibration source:
 \begin{equation}
\mathrm{F_{\rm SIX}\;[\rm erg\;cm^{-2} s^{-1}] = \left(\frac{F_{\rm SIX}\;[cts]}
{F_{\rm Crab}\;[cts]}\right) F_{\rm Crab}\;[erg\;cm^{-2} s^{-1}] },
\end{equation}
where the Crab flux in our survey band is given by 
\begin{equation}
\mathrm{F}_{\rm Crab} = \int_{18 keV}^{55 keV} 
E\;F(E)\:dE .
\label{eq-crab}
\end{equation} 
The Crab spectrum F(E) in units of [photon cm$^{-2}$ sec$^{-1}$ keV$^{-1}$]
is assumed to be power-law shaped having spectral index $\Gamma = -2.15$ and
normalization factor $K = 10.17$.

\subsubsection{Sky coverage}
\begin{figure}[ht!]
\begin{center}
 \includegraphics[scale=0.75]{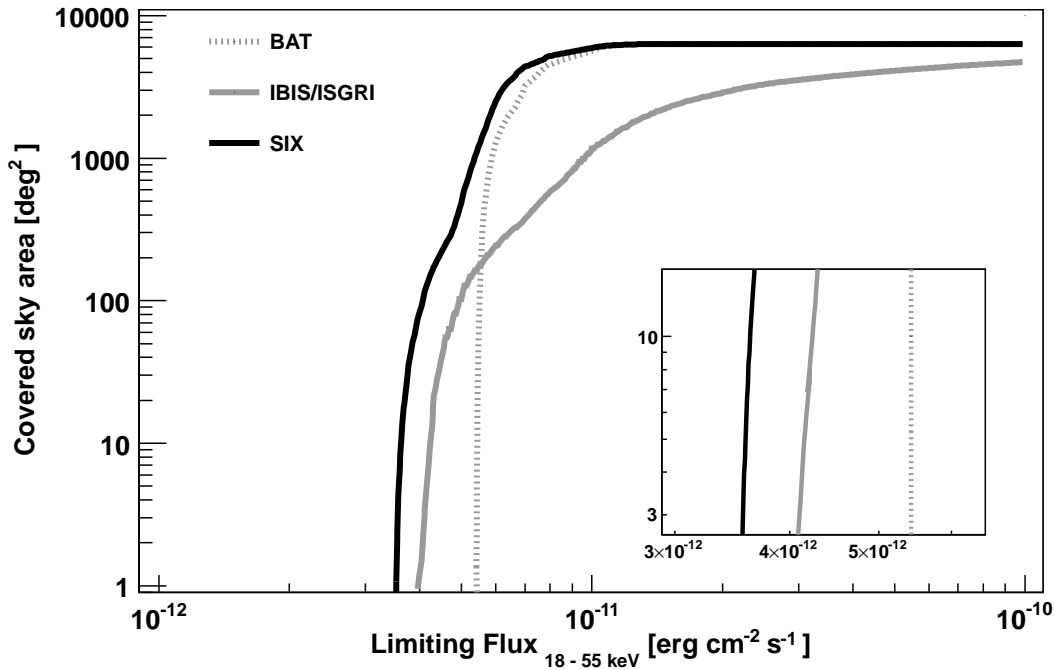}
 \end{center}
\caption{The sky coverages of IBIS/ISGRI (solid gray line), BAT (dotted gray line) 
and {\em SIX} (solid black line). The inset in the lower--right corner is a zoom--in at the level of limiting
fluxes for the 3 surveys. The {\em SIX} sky coverage joins the best of IBIS/ISGRI and BAT
being very uniform  over the whole area and very deep in the center.
\label{fig:skycov}}
\end{figure}

The sky coverage allows one to get a first glance on the uniformity
of the sensitivity over the surveyed sky region. The distribution of the
sky area as function of detection limiting flux is therefore referred to as sky coverage.
The sky coverage as a function of the minimum detectable flux $f_{min}$ is
defined as the sum of the area covered to fluxes $f_i > f_{min}$:
\begin{equation}
\Omega(>f_{min}) = \sum_i^N A_i \ \ \ \ f_i> f_{min} \ \ {\rm [deg^{2}]}
\label{eq:skycov}
\end{equation}
where $A_i$ is the area covered by each pixel and $N$ is the number of
pixels. The minimum detectable flux $f_{min}$ is computed by multiplying the
noise of the area associated to $A_i$ with the detection threshold.\\
As the {\em SIX} survey is the result of 2 independent surveys, we compute and
study all 3 (IBIS/ISGRI, BAT, {\em SIX}) sky coverages. They are plotted in
Figure~\ref{fig:skycov}, where the solid gray line, the dotted gray line and the
solid black  lines are the sky coverages of IBIS/ISGRI, BAT, and {\em SIX} respectively.
It shows that BAT covers the entire surveyed sky area
to a very uniform sensitivity reaching a flux limit of the order of 
$\sim$5 $\times$ 10$^{-12}$ erg cm$^{-2}$ s$^{-1}$. IBIS/ISGRI shows a varying sensitivity
being very deep at the center of its mosaic image (limiting flux
4 $\times$ 10$^{-12}$ erg cm$^{-2}$ s$^{-1}$).

\begin{figure}[ht!]
\begin{center}
 \includegraphics[scale=0.60]{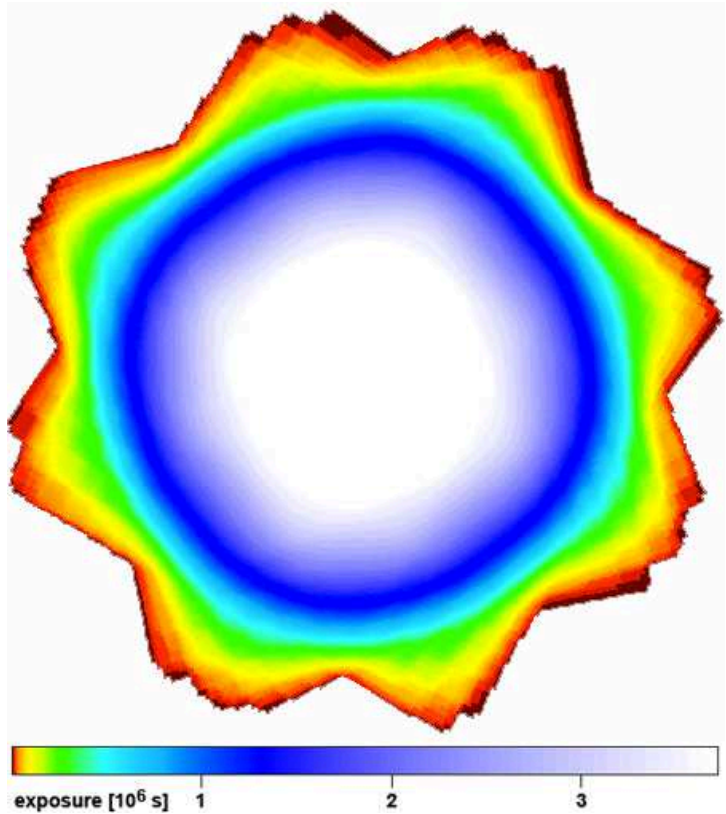}
 \end{center}
\caption{Color coded exposure map of IBIS/ISGRI on the NEP sky area. The innermost
white region is the most exposed sky area corresponding to $\sim$180 deg$^{2}$.
This corresponds in Figure~\ref{fig:skycov} to the value on Y--axis where IBIS/ISGRI
contributes most to the sensitivity limit of the survey. (A color version of this figure is
available in the online journal.)
\label{fig:isgri-nep-exp}}
\end{figure}

The difference between the two performances is mainly related to the
different pointing strategies adopted by the two satellites. 
The main mission objective of {\it Swift} is to study Gamma-Ray Bursts (GRBs).
While chasing up GRBs, {\it Swift}/BAT monitors the sky around the pointing directions.
This permits having a uniform exposure and therefore a uniform sensitivity
over the entire sky.
INTEGRAL is a multiwavelengths observatory. It is the first of this kind. Its
main objective is simultaneously observing objects in gamma--rays, hard and soft
X-rays, and visible light. Therefore, INTEGRAL performs a pointed observing strategy,
where the coordinates of the objects are known in advance. Around these
coordinates INTEGRAL adopts a dither pattern. The dither pattern is a shift
of the center of the instrument FOV with respect to the coordinates of the
object that is being observed.
The pattern consists of a rectangular 5 $\times$ 5 step with angular offset of 2.17$^\circ$
and a small roll angle. As a consequence the innermost
area (a few hundreds deg$^{2}$) of our region of interest is continuously
exposed to IBIS/ISGRI's  fully--coded FOV. This area exhibits the largest
exposure time and therefore it is the sky area exhibiting the best sensitivity
(see the IBIS/ISGRI exposure map in Figure~\ref{fig:isgri-nep-exp}). The extraneous area has
a lower exposure time and it has a lower sensitivity.
The {\em SIX} sky coverage
joins the best of both (IBIS/ISGRI and BAT), being very sensitive and covering the surveyed area very
uniformly at the same time. The whole survey is complete to a flux level
of 10$^{-11}$ erg cm$^{-2}$ s$^{-1}$ while 50\% of the {\em SIX} 
sky is surveyed to 8.5 $\times$ 10$^{-12}$ erg cm$^{-2}$ s$^{-1}$ and the
best flux sensitivity is 3.3 $\times$ 10$^{-12}$ erg cm$^{-2}$ s$^{-1}$.

\subsection{The {\em SIX} catalog}
\begin{figure}[ht!]
\begin{center}
 \includegraphics[scale=0.75]{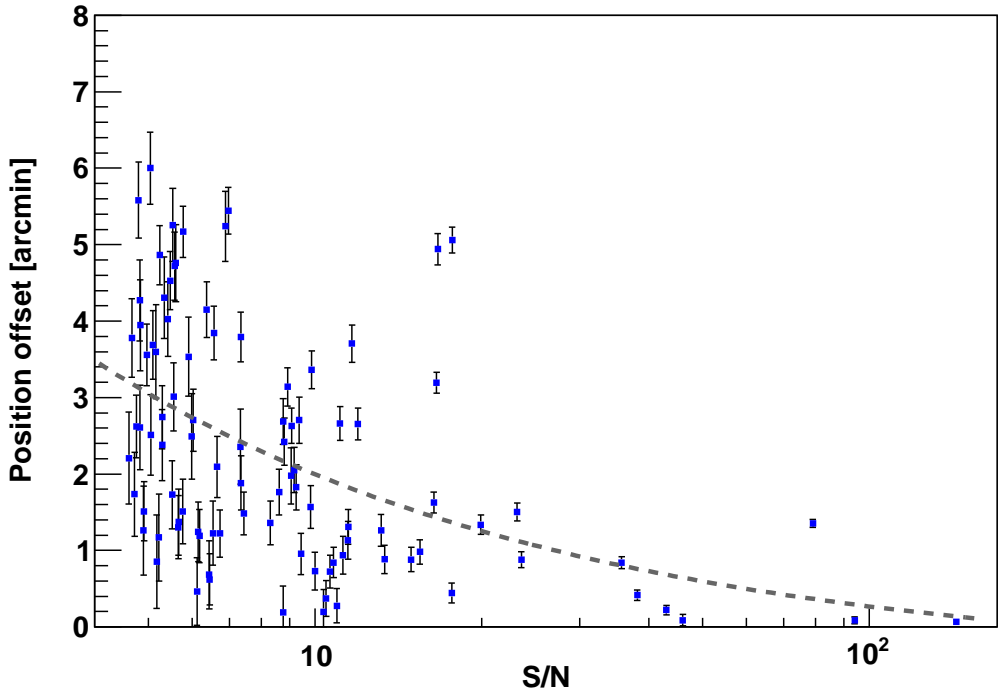}
 \end{center}
\caption{Offset from catalog position of sources detected in the {\em SIX} mosaic
as function of S/N. The dashed line is the function described in
Eq.~\ref{eq:offset}. No sources have an offset larger than 6$\arcmin$.
\label{fig:snr-offset}}
\end{figure}

The {\em SIX} catalog (Table~\ref{tab:sources}) contains 113 sources
having S/N-ratio above 4.8$\sigma$.
To identify this source sample we cross-correlate it with the BAT catalog
\citep{ajello08b,cusumano10}, with the 4$^{th}$ IBIS/ISGRI catalog
\citep{bird10}, and with the INTEGRAL reference catalog
\footnote{http://www.isdc.unige.ch/integral/science/catalogue}. We have
correlated our serendipitously detected objects also with the ROSAT
All-Sky Survey Bright Source Catalogue \citep{voges99}.
In addition, we have made use of the NASA/IPAC Extragalactic Database
(NED)\footnote{http://ned.ipac.caltech.edu/}
and the SIMBAD\footnote{http://simbad.u-strasbg.fr/simbad/}
Astronomical Database. Positional queries were performed considering
possible counterparts within a radius of 6$\arcmin$. 
We searched in literature for the absorption value (N$_{H}$) of each
AGN. When not available, we have derived this
parameter through the soft X-ray spectra. The soft X-ray data come from {\em Chandra},
{\em Swift}/XRT, and {\em XMM-Newton} observations.  {\em Chandra} spectra
were extracted using {\em Chandra} Interactive Analysis of Observations
\citep[CIAO:][]{fruscione06} version 4.4. {\em XMM-Newton} Observation Data Files (ODFs) were
processed using the {\em XMM-Newton} Scientific Analysis Software \citep[SAS:][]{gabriel04}
version 10.0. We used {\em Swift}/XRT data in photon--counting mode only. For the analysis
we used {\texttt{xrtproducts}} and {\em HEAsoft 6.10.2}. These instruments allow us to connect
their spectra to the hard X-ray spectra since their upper energy threshold is between 6--10 keV,
depending on the instrument. The joint fit of the soft X-ray and hard X-ray spectra of
the same source allows derivation of the N$_{H}$ value in excess to the Galactic
column hydrogen density. This latter value was derived using the
database\footnote{http://www.astro.uni-bonn.de/english/tools\_labsearch.php}
accessible on--line and described in \cite{kalberla05}. We used XSPEC 12 \citep{arnaud96}
and the latest available response matrices for calibration to perform the fit.  The best model for the
fit is given by an absorbed power--law with further absorption fixed to the Galactic column
hydrogen density ($wabs(wabs*powerlaw)$). All the other parameters are free to vary. The
N$_{H}$ values and their references, when taken from literature, are reported in
Table~\ref{tab:sources}.\\ 
In addition, the redshifts of our sources are obtained by archive search
of the counterparts. For these identified sources the rest--frame
luminosity was computed in the 18--55 keV energy range using the
equation
\begin{equation}
L_{18-55 keV} = 4 \pi D_L^2 \frac{F_{18-55}}{(1+z)^{2-\Gamma}}
\label{eq-luminosity}
\end{equation}
where $\Gamma$ is the spectral index obtained from the spectral
fit, F$_{18-55 keV}$ is the observed flux in the 18--55 keV energy
range and $D_{L}$ is the luminosity distance.\\
We were able to identify 99 out of the 113 {\em SIX} sources, while 14 are unidentified. Among
these 14 sources 7 are lacking soft X-ray counterparts, while 7 do not have any possible counterpart. 
Table~\ref{tab:source-types} summarizes the types of sources detected in
this survey.
Roughly 16\% of our AGN sample belongs to the blazar subclass.
In their independent surveys IBIS/ISGRI finds 15\% blazars \citep{foschini08} as
does BAT \citep{ajello09b}. These results are in good agreement.
For our blazar sample we have searched
also for counter parts at gamma--ray energies. In order to find spatial coincidences
we have cross--correlated the sky positions of our blazars with the source positions 
of the Second {\it Fermi}--LAT Catalog \citep[2FGL:][]{ackermann11}. To account for the positional
uncertainty of the {\it Fermi}--LAT sources, we find that the sources within 2$\sigma$ confidence
level have an uncertainty $<$ 0.4 deg. Since the positional uncertainty of the {\em SIX} sources
is smaller (6$\arcmin$), we search for spatial coincidences within a conservative error
radius of 0.4 deg. We find that 10 of our blazars coincide with {\it Fermi}-LAT blazars.\\
The identified sources carry the information on the positional accuracy obtained
in our survey. Making use of the sources with known X-ray counterpart, we report in
Figure~\ref{fig:snr-offset} the sources' offset from their catalog positions as functions of
the detection significances. The catalog position is derived from the position of the centroid of
the {\em SIX} PSF that is fit to the region where the sources are detected.
The result is that the {\em SIX} mosaic provides positions accurate to within $\sim$4$\arcmin$
for 95\% of the sample. This is a very good location accuracy.
A fit to the data shows that
the mean offset varies as function of source significance accordingly to:
\begin{equation}
\textup{OFFSET}= (5.95\pm0.28)\times (\textup{S/N})^{-0.38\pm0.06} - (0.42\pm0.29)\ \ \ \ {\rm [arcmin]}
\label{eq:offset}
\end{equation}
A similar dependence on the source significance is known also for IBIS/ISGRI
\citep{gros03,bird06} as well as for BAT \citep{ajello08b,segreto10}. The absence
of very bright sources in the SIX survey does not allow the fit in Figure~\ref{fig:snr-offset}
to be tightly constrained. However, we find that no {\em SIX} source is displaced
by an offset larger than 6$\arcmin$ with respect to the SIMBAD or NED position. Therefore we can
conclude that the {\em SIX} survey locates all sources to better than 6$\arcmin$.\\
As coded--mask detectors have a fairly poor angular resolution, we
consider the possibility of source confusion. IBIS/ISGRI has a narrower PSF
(12$\arcmin$) compared to that of BAT (22$\arcmin$). Resampling the BAT mosaic
image to match the characteristics of the IBIS/ISGRI mosaic image does not
affect the PSF of BAT. Therefore also the angular resolution of BAT is preserved
in the resampled BAT image.
Our new virtual instrument (the combination of BAT and IBIS/ISGRI) has
an angular resolution of 16$\arcmin$ (see Figure~\ref{fig:psf}), which is still
a very good performance.
Taking into account that the total
surveyed sky area is 6200 deg$^2$ we end up with $\sim$48000 possible independent
sky positions for the sources. If our surveyed sky area was covered uniformly to our
limiting flux (3.3 $\times$ 10$^{-12}$ erg cm$^{-2}$ s$^{-1}$) then we would expect
$\sim$1300 sources. Therefore, we can conclude that the source confusion is not an
issue for this survey. In fact, the average source separation of our 113 {\em SIX} sources
is $\sim$7$^{\circ}$ on the 6200 deg$^2$ of sky area.
\begin{table}[ht!]
\begin{center}
\caption{Composition of the {\em SIX} Sample in the 18 -- 55 keV Energy Band.\label{tab:source-types}}
\begin{tabular}{lc}
\hline\hline
Class & Number of Objects  \\
\hline
Seyfert-like AGN & 74 \\
Blazars & 12 \\
Galaxies & 5 \\
Galaxy clusters & 2 \\
Galactic sources & 3 \\
X-ray sources & 3 \\
Unidentified & 14 \\
Total & 113 \\
\hline
\end{tabular}
\end{center}
\end{table}
\clearpage

\subsection{Statistical properties}
\begin{figure}[ht!]
\begin{center}
 \includegraphics[scale=0.75]{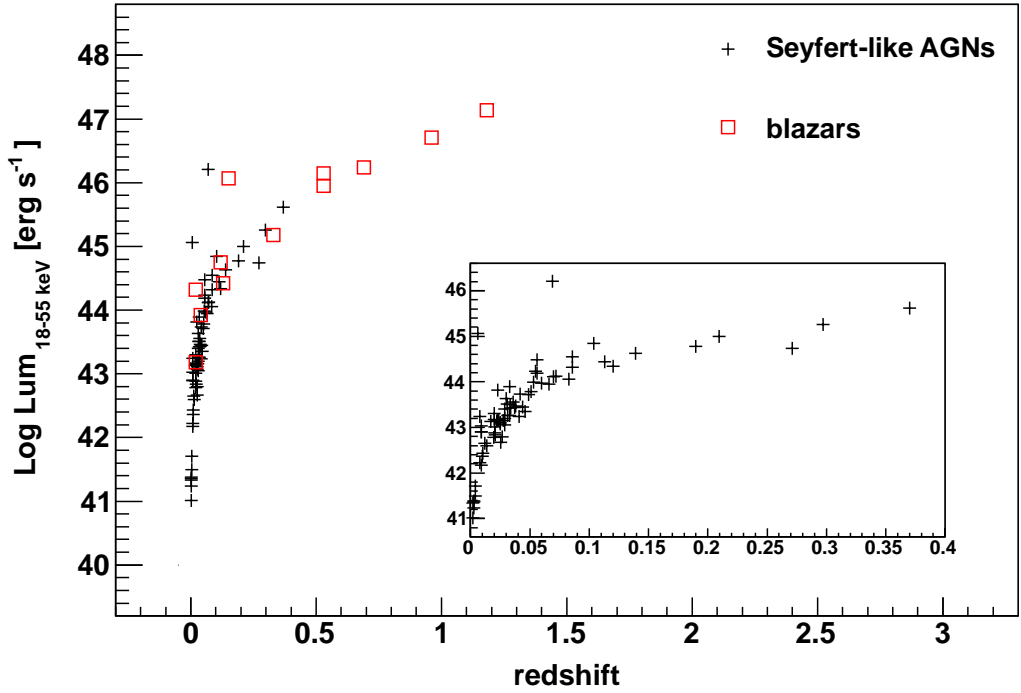}
 \end{center}
\caption{The redshift--luminosity plane shows our flux limited sample split into
different source classes. Crosses are Seyfert--like AGN, while squares
correspond to blazars. These latter sources are detected up to redshift $\sim$1.2. The
inset in the lower right corner is a zoom--in into redshift range 0 $<$ z $<$ 0.4
where only values of Seyfert--like AGN are plotted. (A color version of this figure is
available in the online journal.)
\label{fig:redshift-lum}}
\end{figure}

We use the results of our survey to derive cosmological
information. Figure~\ref{fig:redshift-lum} shows the luminosity-redshift relation
for the identified sources in the 18--55 keV energy band.
Our flux--limited sample shows the clear trend where the most luminous sources are
detected at the greatest distances.
This is of particular importance as the flux--limited AGN sample spans a wide
range in redshift. In Figure~\ref{fig:redshift-lum} black crosses represent
Seyfert--like AGNs and red rectangles are blazars. Seyfert-like AGNs
are sampled within 0 $<$ z $<$ 0.4. For z $>$ 0.4 only blazars are detected.

\subsubsection{Source number--density}
\begin{figure}[ht!]
\epsscale{1.00}
\plotone{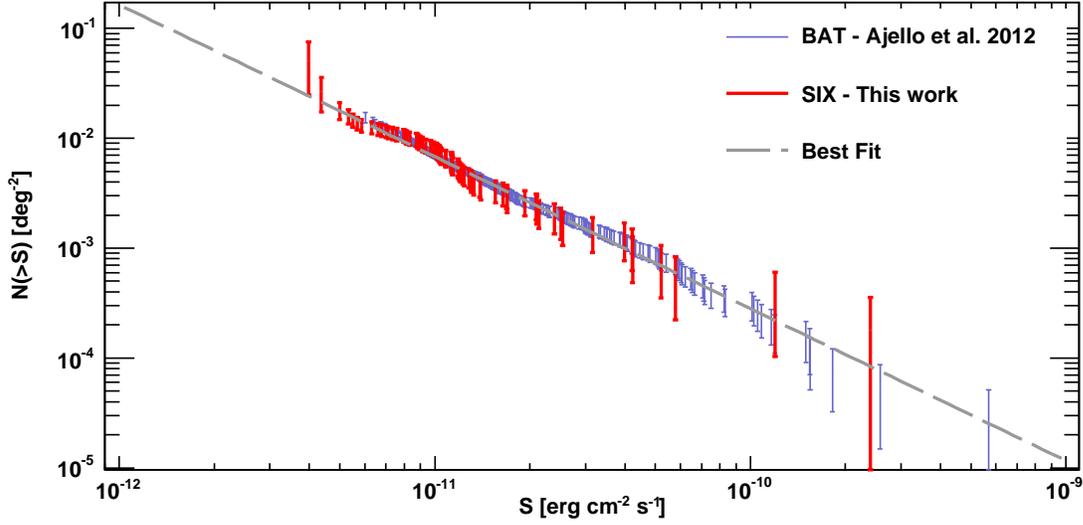}
\caption{The cumulative log $N$--log $S$ distribution for our sample of AGN.
For comparison the 60 months BAT sample \citep{ajello12} is plotted
in background. (A color version of this figure is available in the online journal.)
\label{fig:logn-logs}}
\end{figure}

The log $N$--log $S$ relation represents a tool for detecting a possible cosmological
evolution of a source class. For the {\em SIX} log $N$--log $S$ (see Figure~\ref{fig:logn-logs}) 
we assume a power--law form represented by: N($>$S) = K $\times$ S$^\alpha$ where $N$
is the number of sources above the source flux $S$. The best--fit to the differential
log $N$--log $S$ is expressed by:

\begin{equation}
\mathrm{dN/dS} = (5.62\pm0.65) \times 10^{-18}\;S^{2.38\pm0.11}.
\label{logn-logs}
\end{equation}
 
By integrating the differential function we obtain an
Euclidean slope consistent with a non--evolving population in 
the local Universe. In case of evolution the value of $\alpha$ is expected to be greater than 1.5.
We use the Seyfert--like AGNs reported in \cite{ajello09b} and \cite{krivonos10} as
control samples.

As expected, the power--law slopes agree well. The limiting flux in the {\em SIX}
log $N$--log $S$ is a factor of $\sim$2 fainter and the number--density
of sources is a factor of $\sim$4 higher. In general, the parameters
used to model the number--density functions agree well.
There is general consensus that the peak of the CXB is due to the integrated
emission of unresolved Seyfert--like AGNs \citep{lafranca05, gilli07, treister05, ueda03,
silverman08}. Due to deep observations ({\em Chandra} Deep Field North,
{\em Chandra} Deep Field South and {\em XMM-Newton} Lockman Hole) a
detailed study of the CXB has been performed.
The fraction of intensity due to AGN activity
contributing to the CXB was found to decrease very rapidly with
energy \citep{worsley05}. At its peak ($\sim$30 keV) less than 1\% of the intensity
of the CXB could be attributed to AGN activity \citep{ajello08b}.
The contribution of the {\em SIX}-detected AGNs to the CXB is given by integrating the
number--density $dN$/$dS$ multiplied by the source flux expressed as:

\begin{equation}
\mathrm{F}_{\rm diffuse} = \int_{18 keV}^{55 keV} 
dN/dS\;S\:dS = 1.1356 \times 10^{-9}\; {\rm erg\;cm^{-2}\;s^{-1}\;sr}
\label{eq-diffuse}
\end{equation}

\subsubsection{Luminosity and redshift dependence of the fraction of absorbed AGNs}
\begin{figure}[ht!]
\begin{center}
 \includegraphics[scale=0.35]{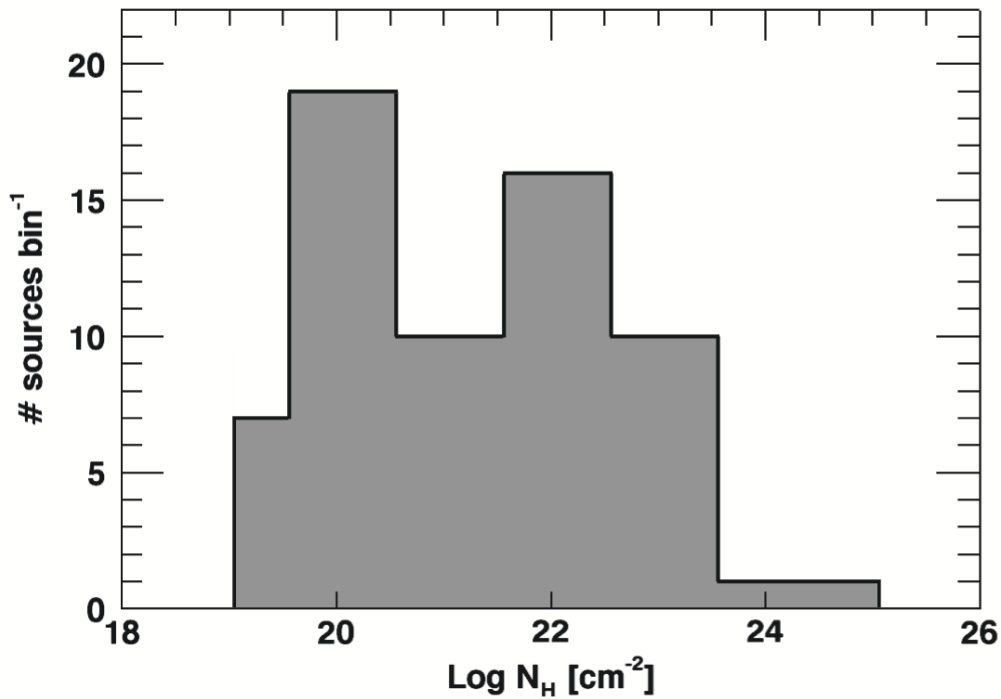}
 \end{center}
\caption{Distribution of intrinsic $N_{H}$--value inferred from
soft X-ray observations. 
\label{fig:nh-frac-abs}}
\end{figure}

As discussed in \S1, the observations in the 18--55 keV energy band are very well
suited for an unbiased (against absorption) detection of AGNs. We have
derived the intrinsic $N_{H}$--value for 64 of our AGNs. In Figure~\ref{fig:nh-frac-abs}
we show the histogram of the observed  $N_{H}$ distribution (in units of number per
bin). As an effective zero value, we set the column densities smaller than log $N_{H}$
$<$ 20 to log $N_{H}$ = 20. The histogram shows that the number of sources drops
for column densities log $N_{H}$ $>$ 23. Roughly $\sim$3\% of our Seyfert--like AGNs
are Compton--thick.\\

\begin{figure}[ht!]
\begin{center}
 \includegraphics[scale=0.80]{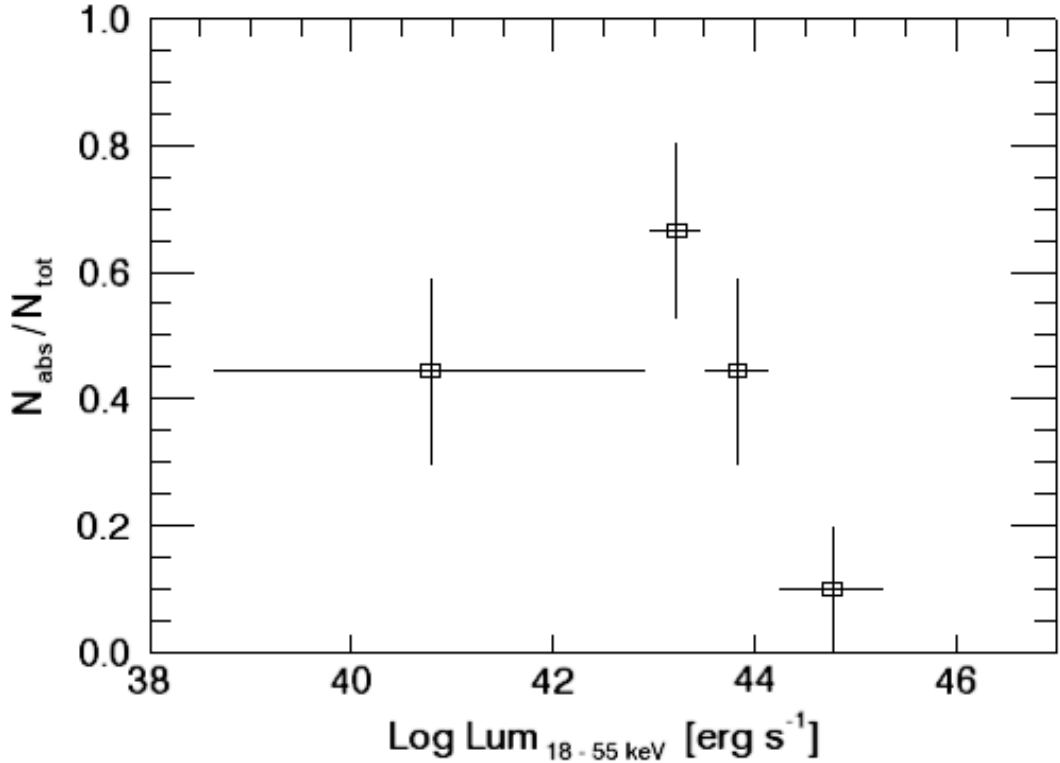}
 \end{center}
\caption{Fraction of obscured AGN (with log N$_{H}$ $>$ 22) vs luminosity. The variables anti-correlate
for luminosities $>$ 10$^{43}$ erg s$^{-1}$.
\label{fig:lum-nh}}
\end{figure}

Our sample covers a wide range in the $L_{X}-z$ plane. Thus, it is possible to
constrain the luminosity dependence and the redshift dependence separately.
Much attention has been paid to an observed anti--correlation of the fraction of
absorbed AGNs as function of luminosity \citep[e.g.][and references therein]{burlon11}.
Such a relation is at odds with the simple AGN unified model \citep{antonucci93,
urry95} since the AGN properties are explained in terms of viewing angle and
no other properties such as luminosity and/or accretion rate are involved. Scenarios
have been proposed where obscuration due to the dust--sublimation radius
depends on the luminosity \citep[$R_{sub} \sim L^{0.5}$:][]{lawrence82} or
where a misaligned disk with respect to the jet axis rules the obscuration \citep{lawrence10}.
But none of them are conclusive. In addition it seems that complex scenarios in the vicinity
of the SMBH do not allow a simple interpretation of the observed anti--correlation.
Considering that our $N_{H}$--inferred AGN sample is incomplete, we
plot the faction of absorbed AGNs vs. luminosity (see Figure~\ref{fig:lum-nh}). 

\begin{figure}[ht!]
\begin{center}
 \includegraphics[scale=0.80]{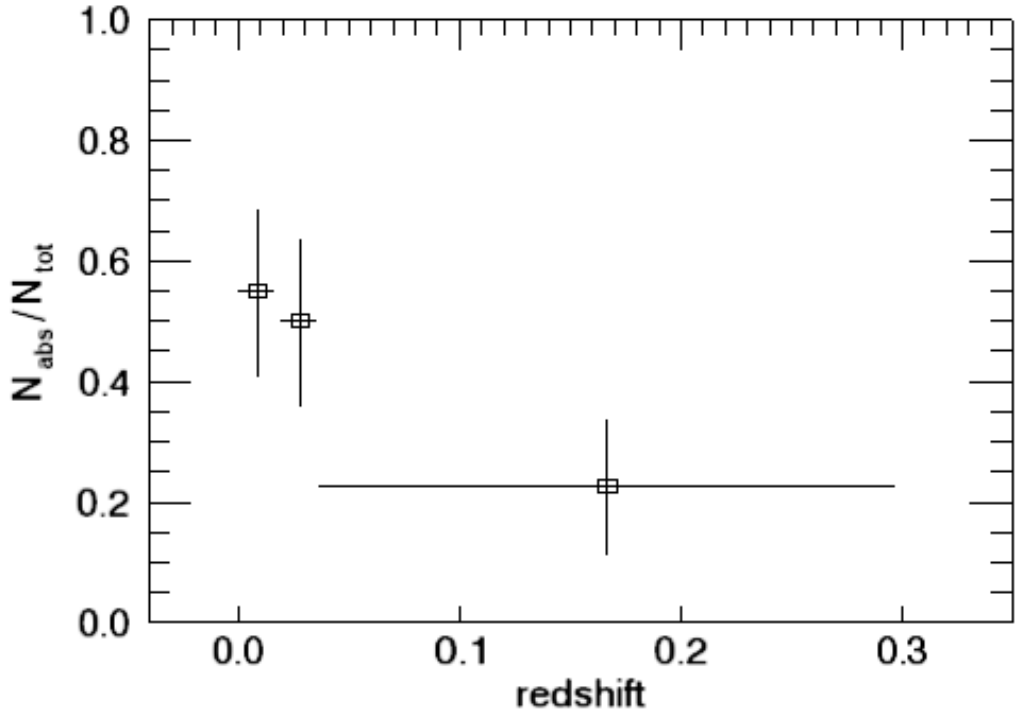}
 \end{center}
\caption{Fraction of obscured AGN (with log N$_{H}$ $>$ 22) vs redshift.
\label{fig:lum-z}}
\end{figure}

The reported errors are calculated as the 1 $\sigma$ binomial confidence interval
as in \cite{gehrels86}. The fraction of the AGNs having log $N_{H}$ $>$ 22 decreases
from 66\% at $L_{X}$=43.2 erg s$^{-1}$ to 10\% at $L_{X}$=44.7 erg s$^{-1}$.
The widths of the luminosity ranges were chosen to contain the same number of sources.
However, to draw conclusions on the shape and the drop of
this relation a better statistic is needed. In this work the low statistics are due to
the limited number of sources. Moreover, 10 AGNs are missing the $N_{H}$
measurement and 14 sources are without any counterpart.\\
Figure~\ref{fig:lum-z} shows the redshift dependence inferred from our sample. The
statistical uncertainties are computed in the same way as for the luminosity
dependence. Even though the uncertainties are rather large, a marginally-significant
decay can be seen from our observations: the fraction of obscured
AGNs declines from 55\% at redshift $z = 0.008$ to 22\% at redshift $z = 0.166$.

\subsubsection{The X-ray luminosity function}
\begin{figure}[ht!]
\begin{center}
 \includegraphics[scale=0.75]{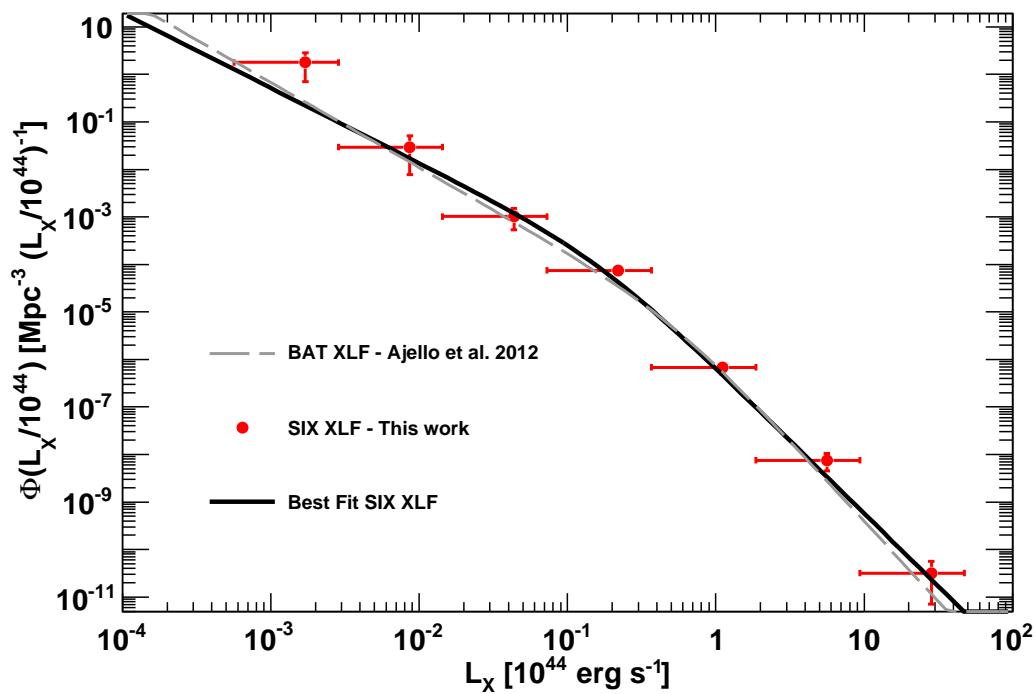}
 \end{center}
\caption{Luminosity function of Seyfert-like AGNs detected in the {\em SIX} survey.
The data (dots with error bars) are represented by a non-evolving double
power-law model (black solid line). Parameters are reported in Table~\ref{tab:xlf}.
For comparison the same model for the BAT-detected (gray dashed line) sources
is plotted. (A color version of this figure is available in the online journal.)
\label{fig:lum-function}}
\end{figure}

The X-ray luminosity functions (XLF) traces distribution and evolution of AGNs throughout the
Universe. In turn this can give hints regarding the formation and growth of
SMBHs and their fueling mechanisms. The {\em SIX} sample is well suited
for this purpose because it has an adequate span in redshift and luminosity.
We consider only those sources from the sample that are identified AGN. 
The AGN evolution and its quantification can be revealed using the 
$V/V_{MAX}$ method proposed in \cite{schmidt68}. Given a flux--limited 
survey and an object with constant luminosity, there is a maximum volume
$V_{max}(z_{max})$ in which the object could have been detected. We
compare $V_{max}(z_{max})$ and the volume in which the object
was effectively detected $V_{z}$. Thus, this latter value can range from 0 to
$V_{max}$. We can compute for each object $i$ the ratio $V_{i}/V_{imax}$.
If the sample is complete and the source number--density constant within
the co--moving volumes then $V_{i}/V_{imax}$ is uniformly distributed
between 0 and 1, implying that the mean value $<$$V/V_{max}$$>$ = 0.5.
Instead if the $<$$V/V_{max}$$>$ $\not=$ 0.5 then
the objects are not uniformly distributed. For $<$$V/V_{max}$$>$ $>$ 0.5
a positive evolution in density or luminosity (or even both) is expected, while
for $<$$V/V_{max}$$>$ $<$ 0.5 a negative evolution of the sample is
expected.
We find that  $<$$V/V_{max}$$>$ = 0.49$\pm$0.02 implying the AGNs
are not evolving in the local Universe. This result is in agreement with those
reported in \cite{tueller08} and in \cite{ajello09b}.\\
The XLF is the co--moving number--density of AGNs having luminosities
[$L,L+dL$]. We represent this function with a double power--law model
as in Equation~\ref{eq:xlf}:
\begin{equation}
\Phi(L_X,z=0) = \frac{d^{2}N}{dV dL_X}=
\frac{K}{ln(10)L_X}\left[ \left(\frac{L_X}{L_*} \right) ^{\gamma_1} 
+ \left(\frac{L_X}{L_*}\right)^{\gamma_2} \right]^{-1}
\label{eq:xlf}
\end{equation}
where $\gamma_{1}$ is the low-end slope, $\gamma_{2}$ is the high-end
slope, $K$ is a normalization factor including the overall density, and $L_{*}$
is the break luminosity. The best-fit parameters are reported in Table~\ref{tab:xlf}.
A visual representation of the XLF in shown in Figure~\ref{fig:lum-function}. The
data points are plotted together with their horizontal and vertical error bars. The
best--fit is obtained by applying the model of Equation~\ref{eq:xlf}. For
comparison we plot the fit result obtained by \cite{ajello12} who used the same
model applied to the BAT sample alone. Furthermore we compare our result with
the results of previous works \citep{beckmann06,sazonov07,tueller08,ajello09b}
performed at hard X-ray energies although not exactly in the same
energy range. The low-end slope, the high-end slope, and the break-luminosity
agree well.

\begin{table}[ht!]
\begin{center}
\caption{XLF in the Energy Range 18--55 keV: Best-fit Parameters\label{tab:xlf}}
\begin{tabular}{cccc}
\hline\hline
K$^{a}$ & $\gamma_1$ & $\gamma_2$ & $L_{*}^{b}$ \\
\hline
6.13$\pm$0.71 & 0.57$\pm$0.08 & 2.07$\pm$0.16 & 0.17$\pm$0.07 \\
\hline
\end{tabular}
\end{center}
$^{a}$in units of 10$^{-5}$ erg$^{-1}$ s Mpc$^{-3}$\\
$^{b}$in units of 10$^{44}$ erg s$^{-1}$
\end{table}
\clearpage

\section{Discussions}
\subsection{The {\em SIX}: the survey of a virtual new X-ray mission}
To capitalize on the advantages of selecting local AGNs at hard X-ray energies we have
combined the observations of IBIS/ISGRI and BAT. This greatly enhances the exposure time
improving the {\em SIX} sensitivity as $t_{exposure}^{-0.5}$.  Moreover,
the systematic uncertainties are minimized because the systematic
errors of both instruments are uncorrelated. This reduces the uncertainties by a
covariance term leading to a limiting flux sensitivity of 3.3 $\times$ 10$^{-12}$ erg
cm$^{-2}$ s$^{-1}$ in the 18--55 keV energy range.\\
BAT contributes to the {\em SIX} survey with a very uniform exposure. The exposure
of IBIS/ISGRI is less uniform (see Figure~\ref{fig:skycov}). This is explained by the different
FOVs of both coded--masks instruments and by the different pointing strategies
that are adopted by the {\it Swift} and the INTEGRAL missions. 
INTEGRAL performs a dithering that leads to a large exposure time
at the center of the surveyed sky area as shown in Figure~\ref{fig:isgri-nep-exp}.
The center of this sky area corresponds roughly to the value of the ordinate
in Figure~\ref{fig:skycov} where the IBIS/ISGRI sky coverage outperforms that
of BAT. IBIS/ISGRI is therefore contributing to the {\em SIX} survey with its long
exposures on limited sky areas.
 \subsection{Results from the sample}
\begin{figure}[ht!]
\begin{center}
 \includegraphics[scale=0.75]{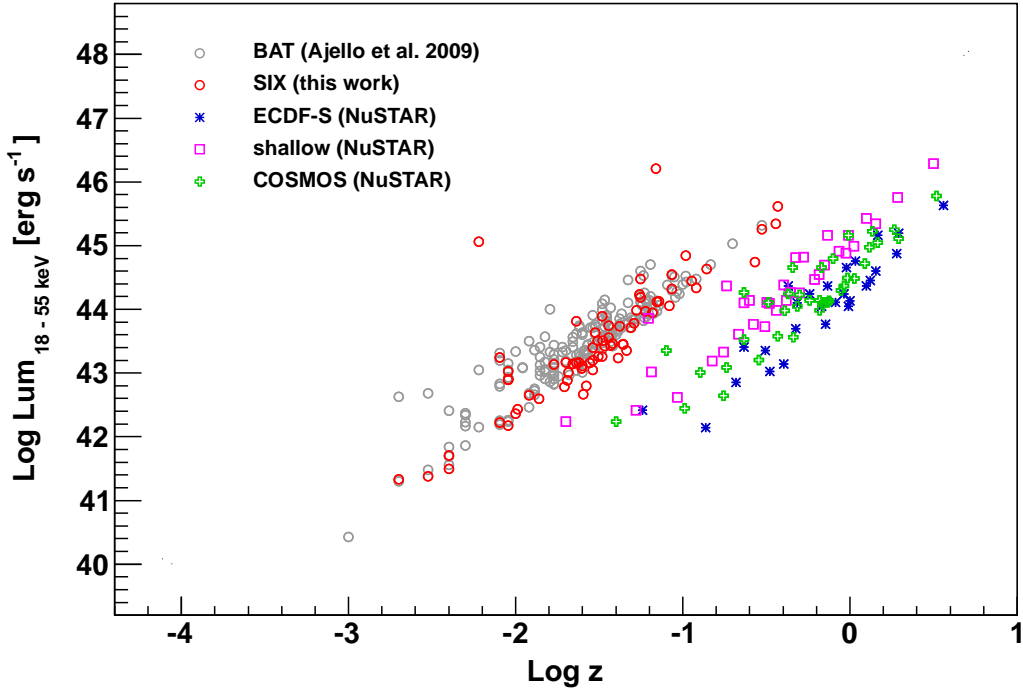}
 \end{center}
\caption{Luminosity-redshift plane of the BAT (gray circles in background) sample,
{\em SIX} sample (red circles in foreground), and predicted NuSTAR samples for different
sky areas \citep{ballantyne11}. (A color version of this figure is available in the online journal.)
\label{fig:z-lum}}
\end{figure}

From observations at soft X--rays we have inferred the $N_{H}$-value
of 64 out of our 74 AGNs. The level of this incompleteness will change as
some of these sources are followed up by {\em Chandra} (PI E. Bottacini
CXC AO-13) and XRT. The fraction of obscured AGNs as function
of luminosity is consistent with a previous work \citep{burlon11}. Also the drop at low
luminosities $\sim$10$^{43}$ erg s$^{-1}$ is reproduced in our work. The relation is
represented by an anti--correlation shown in Figure~\ref{fig:lum-nh} revealing that the $N_{H}$ 
function is luminosity dependent. In contrast to the simple AGN unified scheme \citep{antonucci93,
urry95}, where AGNs have the same geometrical structure irrespective of luminosity and redshift,
this result shows that the opening angle of the torus surrounding the SMBH is larger
for more luminous AGNs. Therefore, high--luminosity AGNs must be able to 'clean out' their
environments. The physical process responsible for this could be the stronger radiation
pressure from the SMBH of the more luminous AGNs. In this scenario the circumnuclear
environment is exposed to higher pressure that causes greater mass
outflow. Indeed, the existence of such mass outflows is supported by recent observations
\citep{pounds09,sturm11,rupke11}. If accompanied by sufficient mechanical energy
transfer, these outflows can provide the coupling of the SMBH and host galaxy co-evolution
\citep{pounds09}.
However, it is ambitious to draw conclusions from the observational fact of the anti--correlation.
How the covering factor of the torus relates to the luminosity depends on the interplay of the SMBH
gravity with the gas, radiation, and magnetic fields accompanied by wind
outflows. The thermal pressure in the gas is able to explain
the low-velocity outflow as X-ray warm absorbers \citep{krolik01}. Instead, the
UV radiatively-driven wind shields itself from X-rays from the central
engine \citep{murray95}. On the other hand, a natural shielding is obtained
in the magneto-centrifugal outflow scenario \citep{everett05}. Finally, in hydrodynamical
simulations \citep{proga04} high-density gas arises naturally that is able to
efficiently absorb X-ray radiation. All these processes show promising results
in numerical simulations.\\
A possible dependence (decrease) of the fraction of absorbed AGNs as function of redshift
is marginally detected in this work (see Figure~\ref{fig:lum-z}). Even though within a larger redshift 
(z $<$ 3) range, a  similar relation is found by \cite{ueda03}. Instead a strong increase of the fraction of
obscured AGN within 0 $<$ z $<$ 2 is found by \cite{lafranca05} and \cite{hasinger08}.
This dependency on the distance at high redshift can be due to the higher gas content
in high--redshift (z $\sim$ 1.5) galaxies \citep{daddi10}. In the local Universe our results
show a mild decay of the fraction of obscured AGNs with redshift.\\
The fit to our flux--number density function is consistent with the Euclidean model.
This is in good agreement with previous measurements by \cite{ krivonos05}, \cite{beckmann06},
\cite{tueller08}, and \cite{ajello09b}. We extend the result towards lower fluxes reaching a flux limit of
a few 10$^{-12}$ cm$^{-2}$ s$^{-1}$. As a result of the increase of the number
density of sources, the {\em SIX} source sample improves the number--density of sources contributing
to the CXB by more than a factor $\sim$~2 compared to the fraction derived from the
sample in \cite{ajello09b}. The contribution of the latter
sample to the CXB is obtained by integrating the emission of the AGN over the
entire extragalactic sky of 30000 deg$^2$. Even though in our work we 
survey $\sim$1/5 of the extragalactic sky, our surveyed sky area is 
particularly suited because of absence of strong sources. 
We can estimate the number of Seyfert--like AGN in a deep NuSTAR \citep{harrison10}
survey. NuSTAR is a NASA mission operating at energies in the range 3--80 keV and
scheduled to be launched in June 2012. It is the first high energy X-ray
mission using focusing optics. Its relatively small FOV ($\sim$ 13 $\times$ 13
arcmin squared) allows surveying small areas compared to coded--mask detectors.
If the luminosity function derived
here does not evolve strongly in either normalization and/or slope
the estimated number of AGN per square degree at energies $\sim$30 keV
is $\sim$100 for a limiting flux of 10$^{-14}$ erg cm$^{-2}$ s$^{-1}$. 
This will be quite possible to be performed by NuSTAR since it will have
an angular resolution $\sim$45 arc sec and source confusion will not
be a problem. 
NuSTAR will perform very sensitive and beam--like extragalactic surveys
tiling not more than a few deg$^{2}$ of sky area.
Detailed predictions \citep{ballantyne11} show, that
independent of NuSTAR's survey strategy (corner shift or half shift) and
the CXB models, AGNs will be sampled most efficiently at redshift z $>$ 0.5.
Figure~\ref{fig:z-lum} displays the redshift - luminosity plane of Seyfert-like
AGNs sampled by BAT \citep[gray circles in background from][]{ajello09b}, {\em SIX}
(solid circles in foreground), and NuSTAR. NuSTAR data are taken from
predictions in \cite{ballantyne11} and adapted to the energy range 18--55 keV.

\begin{figure}[ht!]
\begin{center}
 \includegraphics[scale=0.65]{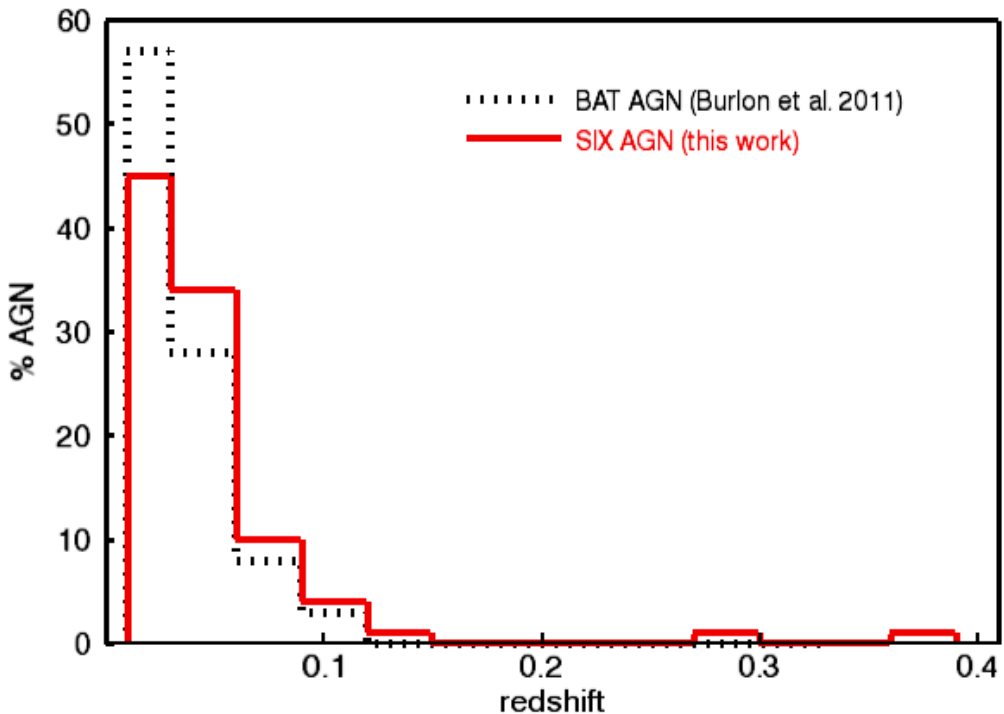}
 \end{center}
\caption{Redshift distribution of the BAT sample (dashed line)
and the {\em SIX} sample (solid line). The {\em SIX} survey samples
the sources systematically at higher redshift compared to the BAT survey.
(A color version of this figure is available in the online journal.)
\label{fig:redshift-distribution}}
\end{figure}

As the NuSTAR survey fields narrow, the redshift distribution of AGNs is shifted
to higher redshift. This is shown in Figure~\ref{fig:z-lum} where squares are from a shallow
survey, crosses from the COSMOS survey (2 deg$^{2}$), and asterisks from the ECDF-S survey
(0.25 deg$^{2}$) assuming an exposure of 6.2 Msec.
NuSTAR's surveys will not compete with the {\em SIX} survey,
but rather they will be complementary. Indeed, if applied to the whole sky the {\em SIX}
survey will fill the redshift and luminosity gap between the current surveys of
IBIS/ISGRI and BAT alone and the NuSTAR surveys. The absorbed {\em SIX} sources
at low redshift are easy follow up targets for NuSTAR.\\
Our small sample used for the XLF does not permit detecting an evolution
in either luminosity or redshift. Therefore, our data are modeled best
by a non-evolving XLF plotted in Figure~\ref{fig:lum-function}.\\
We use the 199 BAT-selected Seyfert-like AGNs of \cite{ajello09b} and later used in \cite{burlon11}
as a control sample in order to evaluate whether the {\em SIX} sample differs in some properties.
For both samples we have derived the redshift distribution excluding radio-loud AGNs and quasars.
The result is plotted in Figure~\ref{fig:redshift-distribution} where the solid
line refers to the {\em SIX} sample and the dashed line represents the sample
of \cite{ajello09b}. Both samples were split into the same bin size. 
About 57\% of the BAT sources are detected within z = 0--0.025. In the same
redshift bin the {\em SIX} survey samples less than 45\% of its sources.

The {\em SIX} survey is more sensitive compared to the BAT survey. Therefore it
samples the sources systematically at higher redshift. Sources at higher redshift
are also the most luminous ones. This is marginally reflected in the XLF
(see Figure~\ref{fig:lum-function})  for luminosities above $10^{45}\;erg\;s^{-1}$.
The best-fit of the XLF adapted from the BAT survey (dashed line) has a slightly steeper
slope at high luminosities. For luminosities below $10^{42}\;erg\;s^{-1}$ a substantial contribution
from the host galaxy to the total luminosity is expected. Therefore, the data point at lowest luminosities
in not properly modeled by the {\em SIX} XLF.\\
Finally, we verify whether the control samples and the {\em SIX} sample are drawn from
the same source population. A Kolmogorov-Smirnov test applied to the two samples with
respect to redshift and luminosity does not allow us to reject the null hypothesis that
the two samples are obtained from the same source population. Therefore, the {\em SIX}
is sampling sources at higher redshift from the same source population that is sampled
by BAT alone. Better statistics for the {\em SIX} XLF can rule out the high-end slope.
This can be achieved by applying the survey to the whole sky.

\section{Conclusions}
A new approach has been developed to survey the sky at hard X-ray energies
(18--55 keV energy band) by combining the observations of {\it Swift}/BAT
and INTEGRAL/IBIS resulting in the {\em SIX} survey.
First we have performed the independent surveys for both instruments. Then we have resampled,
cross-calibrated and merged them. As a result of combining the observations from two different
telescopes, statistical and systematic uncertainties caused by the high background level
of their coded-mask detectors are minimized. In turn the {\em SIX} survey is more sensitive,
like a survey from a virtual new hard X-ray mission.\\
We applied the survey method to 6200 deg$^{2}$ of sky ($\sim$ 20\% of the entire extragalactic sky)
sampling 113 sources that are: 74 Seyfert-like AGNs, 12 blazars, 5 galaxies, 2 clusters of galaxies,
3 Galactic sources, 3 previously detect X-ray sources, and 14 unidentified sources (of which 7 are
newly detected without any counterpart and 7 are of uncertain association).
No false detections due to statistical or systematic fluctuations are expected.
The sources are identified through their soft X-ray counterparts and with {\em Chandra} follow
up observations (CXC AO-12). Unidentified sources are being followed-up in {\em Chandra} AO-13.
Among the AGN sample only two sources are Compton--thick, accounting for $\sim$3\%
of the entire sample.\\
The number density of our identified sources is $\sim$4 times greater than in
our control sample of \cite{ajello09b}. Even though this represents
only a minor fraction of the CXB, the sensitivity improvement with respect to previous measurements
is better than a factor of 2.\\
The fraction of absorbed AGN decreases with increasing luminosities. Although the redshift
dependence is marginally significant, we find a mild decrease of the fraction of obscured AGNs
with increasing redshift. These results require that the covering factor of the torus surrounding the SMBH
changes at least with luminosity.\\
Only robustly identified AGNs were used in our XLF. The data are well represented by
a double power-law model and do not show any evolution in density or in luminosity. The
non-evolving XLF model fits our data best.\\
Based on our results we predict the number density of $\sim$100 Seyfert--like AGNs that
the upcoming NuSTAR mission can detect in 1 deg$^{2}$ of surveyed sky area at a
limiting flux of $10^{-14}\;erg\;cm^{-2}\;s^{-1}$.\\

\acknowledgments
We thank the {\em INTEGRAL} and the {\em Swift} teams for their observations and support.
This research used the NASA/IPAC extragalactic Database (NED), which
is operated by the Jet Propulsion Laboratory, data obtained from the High Energy
Astrophysics Science Archive Research Center (HEASARC) provided by NASA's
Goddard Space Flight Center, and the SIMBAD Astronomical Database, which
is operated by the Centre de Donn\'ees astronomiques de Strasbourg. We also made
use of the TOPCAT tool \citep{taylor05}.
This research has made use of data obtained from the Chandra Data Archive and the
Chandra Source Catalog, and software provided by the Chandra X-ray Center (CXC)
in the application packages CIAO, ChIPS, and Sherpa.
This research is based on observations with {\em INTEGRAL}, an ESA project with instruments and science
data center funded by ESA member states (especially the PI countries: Denmark, France,
Germany, Italy, Switzerland, Spain), and Poland, and with the participation of Russia and
the USA.
We acknowledge G. M. Madejski for very helpful suggestions. We thank S. Digel for comments
on part of the paper. D. R.  Ballantyne is acknowledged for providing predictions on NuSTAR data.
E. B. expresses gratitude to W. Collmar who made it possible to have access to computing
facilities. The authors acknowledge the comments of the anonymous referee, which improved this paper.
This research was started under grant DLR FKZ 50 OR 0405.
E.B. acknowledges support through SAO grants GO1-12144X and GO2-13139X.\\
{\it Facilities:} \facility{Swift (BAT)}, \facility{INTEGRAL (IBIS)}.

\clearpage

\begin{deluxetable}{cclccllccc}
\tablewidth{0pt}
\tabletypesize{\scriptsize}
\rotate
\tablecaption{Detected Hard X-Ray Sources \label{tab:sources}}
\tablehead{
\colhead{R.A.}           & \colhead{Decl.}       &
\colhead{Counterpart}    & \colhead{Flux}      &
\colhead{S/N}            & \colhead{Obj Class} & \colhead{Obj Type} & 
\colhead{Redshift}       & \colhead{Luminosity}  &
\colhead{log $N_H$}\\
\colhead{\scriptsize (J2000)}       & \colhead{\scriptsize (J2000)}  &
\colhead{}                          & \colhead{\scriptsize(10$^{-12}$ erg cm$^{-2}$ s$^{-1}$)} &
\colhead{}                          & \colhead{}         & \colhead{} &
\colhead{\scriptsize(z)}            & \colhead{(erg s$^{-1}$)} &
\colhead{(cm$^{-2}$)}
}
\startdata

133.8152 & 64.4063 & MCG+11-11-032 & 7.97 & 6.44 & AGN & Sy2 & 0.03 & 43.40 & 23.01 [1] \\
148.9211 & 69.6846 & M 82 & 7.09 & 4.90 & galaxy interacting & galaxy interacting & 0.0006 & 39.97 & \nodata \\
149.0083 & 69.0811 & M 81 & 10.1 & 7.34 & AGN & LINER & 0.0001 & 38.64 & 20.53 [1] \\
150.4723 & 55.6943 & 4C 55.19 & 13.9 & 11.4 & AGN & Sy2 & 0.004 & 41.64 & 24.7 [2]\\
161.0304 & 70.4238 & MCG+12-10-067 & 6.31 & 5.67 & AGN & Sy2 & 0.03 & 43.20 & 23.29 [1]\\
163.2331 & 10.6582 & \nodata & 11.6 & 6.67 & \nodata & \nodata & \nodata & \nodata & \nodata \\
166.1311 & 38.2082 & Mrk 421 & 101. & 78.5 & blazar & BL Lac & 0.02 & 44.32 & \nodata \\
166.4922 & 58.9187 & 1RXS J110537.4+585128 & 5.66 & 5.04 & AGN & QSO & 0.19 & 44.77 & 21.20 [1]\\
166.6683 & 72.5697 & NGC 3516 & 57.8 & 46.0 & AGN & Sy1.5 & 0.00 & 42.99 & 21.21 [1]\\
168.9601 & 54.4453 & \nodata & 6.31 & 4.80 & \nodata & \nodata & \nodata & \nodata & \nodata \\
171.3510 & 54.3702 & Mrk 0040 & 10.7 & 9.16 & AGN & Sy1 & 0.02 & 43.01 & 20.90 [2]\\
172.5265 & -14.817 & OM -146 & 17.4 & 8.96 & blazar & FSRQ & 1.18 & 47.14 & \nodata \\
173.1405 & 52.9792 & NGC 3718 & 7.10 & 6.89 & AGN & LINER & 0.003 & 41.23 & 20.00 [2]\\
173.2427 & 10.2765 & 2MASX J11324928+1017473 & 7.57 & 5.99 & AGN & Sy1 & 0.04 & 43.52 & 21.46 [1]\\
174.2112 & 67.6433 & RBS 1004 & 5.50 & 5.42 & blazar & BL Lac & 0.13 & 44.42 & \nodata \\
174.8026 & 59.2078 & RBS 1011 & 11.2 & 9.43 & AGN & Sy1.5 & 0.06 & 43.98 & 19.58 [3]\\
175.5590 & 10.3157 & NGC 3822 & 5.85 & 5.60 & AGN & Sy1 & 0.02 & 42.88 & 20.00 [1]\\
175.9280 & 71.6968 & DO Dra & 15.2 & 10.6 & CV & V* DO Dra & \nodata & \nodata & \nodata \\
176.3579 & 58.9892 & MCG+10-17-061 & 8.39 & 6.15 & galaxy & galaxy & 0.01 & 42.30 & \nodata \\
176.4277 & -18.436 & RBS 1030 & 31.5 & 15.4 & AGN & Sy1 & 0.03 & 43.89 & 20.54 [1]\\
177.0379 & 9.00302 & 2MASX J11475508+0902284 & 8.07 & 5.24 & AGN & Sy? & 0.06 & 43.96 & 21.00 [1]\\
178.0421 & -11.374 & RBS 1044 & 10.3 & 5.52 & AGN & Sy1 & 0.04 & 43.76 & \nodata \\
178.4084 & 49.5092 & RBS 1046 & 4.13 & 4.82 & blazar & FSRQ & 0.33 & 45.18 & \nodata \\
179.5145 & 55.4250 & NGC 3998 & 9.91 & 7.43 & AGN & LINER & 0.003 & 41.46 & 20.09 [1]\\
179.6989 & 42.5570 & IC 751 & 4.36 & 5.90 & AGN & Sy2 & 0.03 & 43.00 & \nodata \\
180.1644 & -1.1668 & 2QZ J120045.2-011041 & 10.1 & 6.73 & AGN & Sy1 & 0.37 & 45.67 & \nodata \\
180.2348 & 6.81394 & CGCG 041-020 & 11.8 & 9.36 & AGN & Sy2 & 0.03 & 43.54 & 22.83 [3]\\
180.2905 & -3.6918 & Mrk 1310 & 5.67 & 5.34 & AGN & Sy1 & 0.01 & 42.69 & 20.72 [1]\\
180.7713 & 44.5210 & NGC 4051 & 25.4 & 23.5 & AGN & Sy1.5 & 0.002 & 41.41 & 20.47 [1]\\
181.5795 & 52.7170 & NGC 4102 & 10.8 & 10.4 & AGN & LINER & 0.002 & 41.29 & 20.94 [1]\\
182.2920 & 47.0460 & Mrk 0198 & 8.73 & 8.30 & AGN & Sy2 & 0.02 & 43.07 & 22.80 [1]\\
182.3597 & 43.6981 & NGC 4138 & 11.3 & 11.2 & AGN & Sy1.9 & 0.002 & 41.33 & 22.90 [3]\\
182.6350 & 39.4063 & NGC 4151 & 239. & 222. & AGN & Sy1 & 0.003 & 42.74 & 22.50 [3]\\
182.6950 & 38.3332 & KUG 1208+386 & 11.8 & 6.88 & AGN & Sy1 & 0.02 & 43.14 & 22.53 [2]\\
183.0916 & -7.5984 & \nodata & 9.56 & 6.69 & \nodata & \nodata & \nodata & \nodata & \nodata \\
183.2574 & 7.03921 & 2MASS J12124981+0659451 & 7.99 & 6.37 & AGN & QSO & 0.20 & 45.00 & \nodata \\
184.2881 & 7.17577 & NGC 4235 & 12.3 & 9.99 & AGN & Sy1 & 0.007 & 42.21 & 21.16 [1]\\
184.7378 & 47.2874 & NGC 4258 & 11.3 & 10.7 & AGN & LINER & 0.001 & 40.77 & 22.91 [4]\\
185.5332 & 75.3006 & Mrk 205 & 11.9 & 8.60 & AGN & Sy1 & 0.07 & 44.15 & 20.88 [5]\\
185.6009 & 4.20212 & 4C 04.42 & 10.5 & 9.81 & blazar & FSRQ & 0.96 & 46.70 & \nodata \\
185.8403 & 2.68141 & Mrk 50 & 12.8 & 8.75 & AGN & Sy1 & 0.02 & 43.19 & 20.92 [1]\\
186.4482 & 12.6643 & NGC 4388 & 119. & 94.0 & AGN & Sy2 & 0.008 & 43.27 & 23.63 [3]\\
187.2817 & 2.04932 & 3C 273 & 172. & 143. & blazar & FSRQ & 0.15 & 46.07 & \nodata \\
188.6686 & 52.6444 & \nodata & 6.42 & 5.11 & \nodata & \nodata & \nodata & \nodata & \nodata \\
189.6855 & 9.46984 & 2MASX J12384342+0927362 & 6.61 & 5.66 & AGN & Sy2 & 0.08 & 44.05 & \nodata \\
189.7676 & -16.184 & IGR J12391-1612 & 16.8 & 10.9 & AGN & Sy2 & 0.03 & 43.72 & 22.48 [1]\\
189.9053 & -5.3471 & NGC 4593 & 42.2 & 38.1 & AGN & Sy1 & 0.009 & 42.88 & 20.30 [3]\\
191.7000 & 54.5375 & NGC 4686 & 9.72 & 9.24 & AGN & Sy2 & 0.01 & 42.78 & 23.84 [1]\\
192.8286 & -11.722 & \nodata & 9.22 & 5.81 & \nodata & \nodata & \nodata & \nodata & \nodata \\
193.0649 & -13.419 & NGC 4748 & 9.34 & 6.18 & AGN & Sy1 & 0.01 & 42.59 & 20.77 [1]\\
193.9939 & 4.33340 & \nodata & 5.84 & 5.21 & \nodata & \nodata & \nodata & \nodata & \nodata \\
194.0513 & -5.7909 & 3C 279 & 12.4 & 10.3 & blazar & FSRQ & 0.53 & 46.15 & \nodata \\
195.9917 & 53.7738 & IGR J13038+5348 & 16.4 & 14.9 & AGN & Sy1 & 0.02 & 43.52 & 20.81 [1]\\
196.0532 & -5.5644 & NGC 4941 & 9.11 & 6.45 & AGN & Sy2 & 0.003 & 41.43 & 22.95 [1]\\
196.0877 & -10.309 & NGC 4939 & 12.0 & 8.75 & AGN & Sy2 & 0.01 & 42.46 & \nodata \\
196.9632 & -2.0556 & \nodata & 7.63 & 7.48 & \nodata & \nodata & \nodata & \nodata & \nodata \\
197.2785 & 11.6407 & NGC 4992 & 20.9 & 17.6 & AGN & Sy2 & 0.02 & 43.48 & 23.74 [6]\\
198.2996 & -11.127 & RBS 1233 & 8.95 & 6.65 & AGN & Sy1 & 0.03 & 43.38 & 20.74 [1]\\
198.8493 & 44.4093 & Mrk 248 & 11.3 & 11.4 & AGN & Sy2 & 0.03 & 43.51 & 22.81 [2]\\
199.7661 & -9.3549 & \nodata & 6.85 & 5.89 & \nodata & \nodata & \nodata & \nodata & \nodata \\
200.2616 & 8.96113 & NGC5100 & 6.13 & 5.29 & galaxy group & galaxy group & 0.03 & 43.15 & \nodata \\
200.6131 & -16.733 & MCG-03-34-063 & 21.3 & 11.0 & AGN & Sy2 & 0.01 & 43.12 & 23.59 [3]\\
202.1334 & -1.5129 & \nodata & 6.34 & 5.36 & \nodata & \nodata & \nodata & \nodata & \nodata \\
203.6978 & -23.425 & ESO 509-66 & 13.2 & 5.77 & AGN & Sy2 & 0.04 & 43.79 & 23.05 [1]\\
203.9518 & 3.01973 & NGC 5231 & 8.80 & 5.29 & AGN & \nodata & 0.02 & 42.97 & 22.23 [1]\\
204.3828 & -13.032 & QSO B1334-127 & 7.91 & 4.83 & blazar & BL Lac & 0.53 & 45.95 & \nodata \\
204.5677 & 4.54615 & NGC 5252 & 52.1 & 43.0 & AGN & Sy2 & 0.02 & 43.76 & 22.34 [8]\\
205.0111 & 55.8247 & \nodata & 8.60 & 5.48 & \nodata & \nodata & \nodata & \nodata & \nodata \\
205.3109 & -14.660 & RBS 1303 & 13.0 & 6.54 & AGN & Sy1 & 0.04 & 43.72 & 21.23 [1]\\
206.3981 & 41.6618 & NGC 5290 & 9.73 & 5.77 & galaxy group & galaxy group & 0.00 & 42.20 & \nodata \\
208.0023 & -18.300 & \nodata & 16.9 & 6.12 & \nodata & \nodata & \nodata & \nodata & \nodata \\
208.3393 & 69.3013 & Mrk 279 & 21.0 & 16.3 & AGN & Sy1 & 0.03 & 43.65 & 20.53 [3]\\
208.4456 & -11.406 & \nodata & 7.69 & 4.80 & \nodata & \nodata & \nodata & \nodata & \nodata \\
209.0421 & 38.5687 & Mrk 0464 & 10.2 & 8.91 & AGN & Sy1 & 0.05 & 43.79 & 20.00 [2]\\
213.3866 & -3.2043 & NGC 5506 & 131. & 79.0 & AGN & Sy1.9 & 0.00 & 43.03 & 22.53 [3]\\
215.4348 & 47.7875 & RBS 1378 & 10.4 & 9.36 & AGN & Sy1 & 0.07 & 44.11 & 21.26 [2]\\
216.5659 & 37.8241 & ABELL 1914 & 5.11 & 5.16 & galaxy cluster & galaxy cluster & 0.17 & 44.61 & \nodata \\
217.2120 & 42.6515 & H 1426+428 & 13.0 & 11.6 & blazar & BL Lac & 0.12 & 44.75 & \nodata \\
217.3623 & 1.31454 & Mrk 1383 & 12.2 & 6.97 & AGN & Sy1 & 0.08 & 44.35 & 20.00 [2]\\
218.4907 & 5.47754 & NGC 5674 & 8.13 & 5.55 & AGN & Sy1 & 0.02 & 43.06 & \nodata \\
218.8075 & 48.6441 & NGC 5683 & 6.77 & 5.09 & AGN & Sy1 & 0.04 & 43.42 & \nodata \\
219.1769 & 58.7837 & Mrk 817 & 15.5 & 11.9 & AGN & Sy1 & 0.03 & 43.54 & 23.49 [2]\\
219.3705 & 58.9051 & \nodata & 8.57 & 7.18 & \nodata & \nodata & \nodata & \nodata & \nodata \\
220.2547 & 53.4781 & Mrk 477 & 9.68 & 7.34 & AGN & Sy2 & 0.03 & 43.51 & 24.00 [7]\\
220.6846 & -17.225 & NGC 5728 & 42.2 & 17.6 & AGN & Sy2 & 0.01 & 42.92 & 24.14 [9]\\
228.8484 & 42.0416 & NGC 5899 & 8.82 & 6.56 & AGN & Sy2 & 0.01 & 42.15 & 23.12 [2]\\
229.8817 & 65.6329 & MCG+11-19-006 & 7.33 & 4.91 & AGN & Sy2 & 0.04 & 43.52 & 21.38 [1]\\
234.0596 & 57.8813 & Mrk 290 & 13.0 & 9.86 & AGN & Sy1 & 0.02 & 43.41 & 20.40 [3]\\
236.6410 & 69.4657 & 2MASX J15462424+6929102 & 5.66 & 5.17 & galaxy & galaxy & \nodata & \nodata & \nodata \\
243.5509 & 65.6973 & Mrk 876 & 8.32 & 6.02 & AGN & Sy1 & 0.11 & 44.49 & 19.06 [1]\\
245.0576 & 81.0390 & MCG+14-08-004 & 10.4 & 8.79 & AGN & Sy2 & 0.02 & 43.13 & 22.88 [1]\\
247.1031 & 51.7736 & Mrk 1498 & 23.9 & 16.5 & AGN & Sy1.9 & 0.05 & 44.24 & 23.26 [3]\\
253.1046 & 55.9419 & MCG+09-28-001 & 5.49 & 4.83 & AGN & Sy2 & 0.02 & 43.02 & 22.66 [1]\\
259.9293 & 48.9820 & Arp 102B & 10.7 & 5.47 & blazar & FSRQ & 0.02 & 43.18 & \nodata \\
260.5280 & 43.2471 & FIRST J172201.9+431523 & 9.15 & 5.05 & AGN & Sy1 & 0.13 & 44.67 & 21.12 [1]\\
270.0433 & 66.5841 & NGC 6552 & 4.98 & 4.96 & AGN & Sy2 & 0.02 & 42.90 & \nodata \\
274.0939 & 49.8605 & AM Her & 35.0 & 23.1 & CV & CV & \nodata & \nodata & \nodata \\
275.5807 & 64.3694 & 1ES 1821+643 & 7.00 & 9.06 & AGN & Sy1 & 0.29 & 45.29 & 20.00 [2]\\
277.5011 & 48.7652 & 3C 380 & 8.34 & 5.58 & blazar & \nodata & 0.69 & 46.24 & \nodata \\
280.6068 & 79.7714 & 3C 390.3 & 39.8 & 35.7 & AGN & Sy1 & 0.05 & 44.47 & 21.03 [3]\\
281.3777 & 72.1933 & 2MASX J18452628+7211008 & 3.97 & 5.22 & AGN & Sy2 & 0.04 & 43.29 & 22.82 [1]\\
290.2788 & 43.9628 & ACO 2319 & 24.2 & 16.6 & galaxy cluster & galaxy cluster & 0.05 & 44.25 & \nodata \\
291.1899 & 50.2373 & CH Cyg & 11.8 & 7.33 & symbiotic star & symbiotic star & \nodata & \nodata & \nodata \\
291.2818 & 50.6906 & 2E~1923.7+5037 & 7.98 & 4.92 & X-ray source & X-ray source & \nodata & \nodata & \nodata \\
291.7341 & 41.5856 & 1RXS J192630.6+413314 & 10.8 & 5.53 & X-ray source & X-ray source & \nodata & \nodata & \nodata \\
292.1210 & 73.9471 & 1ES 1928+73.8 & 5.32 & 4.98 & AGN & Sy1 & 0.03 & 43.20 & 20.93 [1]\\
296.8676 & 44.8043 & CXOU J194719.3+444942 & 13.8 & 9.04 & AGN & Sy2 & 0.05 & 43.96 & \nodata \\
299.7378 & 40.8185 & CXO J195857.9+404856 & 38.2 & 31.4 & X-ray source & X-ray source & \nodata & \nodata & \nodata \\
300.0223 & 65.1576 & 1ES 1959+650 & 15.5 & 13.3 & blazar & BL Lac & 0.04 & 43.92 & \nodata \\
310.7268 & 75.1344 & 4C 74.26 & 25.0 & 19.9 & AGN & Sy1 & 0.10 & 44.83 & 21.22 [1]\\
311.7313 & 65.2038 & \nodata & 5.67 & 4.81 & \nodata & \nodata & \nodata & \nodata & \nodata \\
318.6292 & 82.0649 & S5 2116+81 & 19.2 & 13.1 & AGN & Sy1 & 0.08 & 44.54 & 21.03 [1]\\
337.3319 & 66.7884 & IGR J22292+6647 & 11.2 & 6.12 & AGN & Sy? & 0.11 & 44.56 & 21.77 [1]\\
\enddata
\tablerefs{
[1] this work;
[2] \citealp{burlon11};
[3] \citealp{tueller08};
[4] \citealp{cappi06};
[5] \citealp{page05};
[6] \citealp{winter08};
[7] \citealp{shu07};
[8] \citealp{dadina10};
[9] \citealp{comastri10}.
}
\end{deluxetable}
\clearpage

\end{document}